\documentclass[useAMS,usenatbib,galley]{mn2e}
\def\spose#1{\hbox to 0pt{#1\hss}}
\def\approxlt{\mathrel{\spose{\lower 3pt\hbox{$\sim$}}
	\raise 2.0pt\hbox{$<$}}}
\def\approxgt{\mathrel{\spose{\lower 3pt\hbox{$\sim$}}
	\raise 2.0pt\hbox{$>$}}}
\def\approxpropto{\mathrel{\spose{\lower 3pt\hbox{$\sim$}}
	\raise 2.0pt\hbox{$\propto$}}}
\mathchardef\twiddle="2218

\def\multleft#1{\hbox to size{\vbox {\halign {\lft{##}\cr #1}}\hfill}\par}
\def\multright#1{\hbox to size{\vbox {\halign {\rt{##}\cr #1}}\hfill}\par}

\def\today{\ifcase\month\or January\or February\or March\or April\or May\or
      June\or July\or August\or September\or October\or November\or December\fi
      \space\number\day, \number\year}
\def\<{\thinspace}

\def\arcsec{{\rm\thinspace arcsec}}
\def\cm{{\rm\thinspace cm}}
\def\erg{{\rm\thinspace erg}}

\def\keV{{\rm\thinspace keV}}

\def\km{{\rm\thinspace km}}
\def\kpc{{\rm\thinspace kpc}}

\def\Mpc{{\rm\thinspace Mpc}}

\def\pc{{\rm\thinspace pc}}

\def\s{{\rm\thinspace s}}

\def\ergpcmsqps{\hbox{$\erg\cm^{-2}\s^{-1}\,$}}

\def\ergps{\hbox{$\erg\s^{-1}\,$}}

\usepackage{graphicx}
\usepackage{txfonts}
\usepackage{natbib}
\usepackage{wrapfig}                  
\usepackage{amssymb}
\usepackage{longtable}
\usepackage{times}
\usepackage[usenames]{color}
\usepackage{psfig}
\usepackage{soul,color}
\usepackage{subfigure}
\newcommand*{\mysub}[2]{\ensuremath{#1_{\mathrm{#2}}}}
\newcommand*{\Omegam}{\mysub{\Omega}{m}}

\newcommand*{\Omegal}{\ensuremath{\Omega_{\Lambda}}}

\newcommand*{\LCDM}{\ensuremath{\Lambda}CDM}

\newcommand*{\msolar}{\mysub{M}{\odot}}

\newcommand*{\thresh}{\mysub{P}{b} \ensuremath{< 10^{-3}}}

\title[The Triggering Mechanims for Cluster AGN]
  { X-ray bright active galactic nuclei in massive galaxy clusters III: New insights into the triggering mechanisms of cluster AGN }
\author[S. Ehlert et al.]
{S.~Ehlert,$^{1,2,3,4}$\thanks{Email:sehlert@space.mit.edu.}
S.W.~Allen,$^{1,2,3}$
W.N.~Brandt,$^{6,7}$
R.E.A.~Canning,$^{1,2,3}$
\newauthor
B.~Luo,$^{6,7}$
A.~Mantz,$^{9,10}$
R.G.~Morris,$^{1,2,3}$
A.~von der Linden,$^{1,3,5}$ 
and Y.Q.~Xue,$^{8}$\\
$^1$Kavli Institute for Astrophysics and Space Research, Massachusetts Institute of Technology, 77 Massachusetts Ave., Cambridge, MA 02139, USA\\
$^2$SLAC National Accelerator Laboratory, 2575 Sand Hill Road, Menlo Park, CA 94025, USA\\
$^3$Department of Physics, Stanford University, 382 Via Pueblo Mall, Stanford, CA 94305-4060, USA\\
$^4$Kavli Institute for Particle Astrophysics and Cosmology, 452 Lomita Mall, Stanford, CA 94305-4085, USA\\
$^5$Dark Cosmology Centre, Niels Bohr Institute, University of Copenhagen, Juliane Maries Vej 30, 2100 Copenhagen, Denmark\\
$^6$ Department of Astronomy and Astrophysics, Pennsylvania State University, University Park, PA 16802, USA\\
$^7$ Institute for Gravitation and the Cosmos, Department of Physics, Pennsylvania State University, University Park, PA 16802, USA\\
$^8$ Key Laboratory for Research in Galaxies and Cosmology, Department of Astronomy, \\
University of Science and Technology of China, Chinese Academy of Sciences, Hefei, Anhui 230026, China\\
$^9$ Kavli Institute for Cosmological Physics, 5640 S. Ellis Avenue, Chicago, IL 60637, USA\\
$^{10}$ Department of Astronomy and Astrophysics, University of Chicago, 5640 S. Ellis Avenue, Chicago, IL 60637, USA}

\def\cha{{\it Chandra}}
\def\ae{{\small ACIS-EXTRACT}}

\def\rfive{\mysub{r}{500}}
\def\rosat{{\it ROSAT}}

\def\sext{{\small SEXTRACTOR}}

\def\arcsec {\hbox{$^{\prime\prime}$}}

\def\cdfs{{\it CDFS}}

\begin{document}

\maketitle

\begin{abstract}
We present the results of a new analysis of the X-ray selected Active Galactic Nuclei (AGN) population in the vicinity of 135 of the most massive galaxy clusters in the redshift range of $0.2 < z < 0.9$ observed with \cha. With a sample of more than 11,000 X-ray point sources, we are able to measure, for the first time, evidence for evolution in the cluster AGN population beyond the expected evolution of field AGN. Our analysis shows that overall number density of cluster AGN scales with the cluster mass as $\sim \mysub{M}{500}^{-1.2}$. There is no evidence for the overall number density of cluster member X-ray AGN depending on the cluster redshift in a manner different than field AGN, nor there is any evidence that the spatial distribution of cluster AGN (given in units of the cluster overdensity radius \rfive) strongly depends on the cluster mass or redshift. The $M^{-1.2 \pm 0.7}$ scaling relation we measure is consistent with theoretical predictions of the galaxy merger rate in clusters, which is expected to scale with the cluster velocity dispersion, $\sigma$, as $\sim \sigma^{-3}$ or $ \sim M^{-1}$. This consistency suggests that AGN in clusters may be predominantly triggered by galaxy mergers, a result that is further corroborated by visual inspection of {\it Hubble} images for 23 spectroscopically confirmed cluster member AGN in our sample. A merger-driven scenario for the triggering of X-ray AGN is not strongly favored by studies of field galaxies, however, suggesting that different mechanisms may be primarily responsible for the triggering of cluster and field X-ray AGN. 
   
\end{abstract}

\section{Introduction}

Modern studies of galaxy evolution have demonstrated the essential role central supermassive black holes can play in establishing the observed properties of galaxies on both small ($\sim 1 \pc$) and large ($\sim 1 \kpc$) scales \citep[e.g.][]{Silk1998,Ferrarese2000}. The powerful outbursts of Active Galactic Nuclei (AGN) appear to be an especially important component of modern galaxy formation models, but many aspects of the connection between galaxies and their AGN remain largely uncertain.  
 
Luminous AGN emit across the entire electromagnetic spectrum, but point-like X-ray emission is one of the cleanest signatures of AGN \citep{Brandt2005} making X-ray observations particularly well suited for AGN population studies. Deep field surveys of X-ray AGN have already made great progress in understanding the typical host galaxies and environments of AGN. X-ray selected AGN are predominantly hosted in the most massive galaxies, with stellar masses ($\mysub{M}{\star}$) of $\mysub{M}{\star} \gtrsim 10^{10} \msolar$ \citep{Xue2010}. Comparing the host galaxies of X-ray AGN on the color-magnitude diagram shows that they occupy the same locus of this phase space as normal galaxies with similar stellar masses, suggesting that the host galaxy mass plays the key role in determining the likelihood of it hosting an AGN. This is further supported by investigations into the clustering of X-ray AGN in the field \citep[see][for a review]{Cappelluti2012}, which indicate that the majority of X-ray AGN are hosted in dark matter halos with masses of $M \sim 10^{13} \msolar h^{-1}$, a value that is consistently measured out to redshifts of $z \sim 2$. 

Although it is clear from these studies that the typical host galaxy environment for an AGN is reasonably well constrained to be self-similar out to redshifts of $z \sim 2$, the triggering mechanism or mechanisms that cause these luminous AGN outbursts remain largely uncertain. Deep field, multiwavelength surveys of X-ray AGN suggest that different mechanisms may dominate for galaxies at different redshifts \citep[see][for a review]{Cappelluti2012}. Major galaxy mergers are likely the most important mechanisms for fueling quasars at the highest luminosities and redshifts \citep[$L \gtrsim 10^{44} \ergps$, e.g.][]{Hopkins2006,Hopkins2008,Hasinger2008}. At lower redshifts ($z \lesssim 1$), bar instabilities and less extreme galaxy-galaxy interactions are inferred to be more efficient at producing AGN \citep[e.g.][]{Georgakakis2009}. Investigations into the properties of the galaxies hosting AGN indicate that their morphologies are similar to comparable galaxies that do not host AGN \citep[e.g.][]{Reichard2009,Tal2009}.

One useful way to explore these triggering mechanisms is to observe the AGN populations in massive galaxy clusters. Galaxy clusters are not only sites of large numbers of galaxies in close proximity to one another but also host a hot, diffuse, X-ray bright intracluster medium (ICM) \citep[e.g.][]{Sarazin1988}. Both factors are expected to play a role in transforming galaxies in clusters, through tidal encounters, mergers between neighboring galaxies \citep{Mamon1992,Moore1998}, or by galaxy-ICM interactions such as ram pressure stripping \citep[e.g.][]{Gunn1972}. Studying how the AGN population in clusters is related to the host cluster properties allows us to understand more completely how the variations in the merger frequency or density of the ICM may influence a galaxy's ability to host an AGN outburst. 

Previous studies have established that galaxies in local clusters have lower average star formation rates than the field \citep[e.g.][]{Dressler1980}. Previous studies of the X-ray AGN population in galaxy clusters, however, have typically suffered from limited source statistics. Because the fraction of galaxies hosting X-ray AGN is typically of order $\sim 0.1-1\%$ \citep[e.g.][]{Haggard2010}, large samples of galaxy clusters are required to measure the cluster-specific AGN population with high precision. Understanding how the AGN population varies with cluster mass and redshift additionally requires detailed spectroscopy and mass proxy information that is only just becoming available \citep{Mantz2010a,Mantz2010b,VonderLinden2012,Kelly2012,Applegate2012}. Finally, any attempt to measure the cluster-specific influences on their constituent AGN population must also account for the cosmic evolution of X-ray AGN in the field (also known as the X-ray Luminosity Function or XLF) which has already been measured to have a strong redshift dependence \citep[e.g.][]{Ueda2003,Hasinger2005,Ueda2014}. 

In this paper, we expand the analysis of \cite[hereafter Paper I and Paper II, respectively]{Ehlert2013,EhlertFrac} to a larger sample of galaxy clusters to of test for the presence of a cluster mass and/or redshift dependent signal beyond those expected from field evolution. With more than 11,000 X-ray AGN cataloged here we are able to, for the first time, quantify the extent to which the X-ray AGN population in galaxy clusters may depend on the host cluster mass and redshift. The presence or absence of these signals offers important new evidence as to the key astrophysical processes that drive the evolution of AGN in clusters. When calculating distances, we assume a \LCDM \ cosmological model with \Omegam=0.3, \Omegal=0.7, and $H_{0}=70 \km \s^{-1} \Mpc^{-1}$.

\section{The Cluster Sample}

 The clusters included in our study have been drawn from wide-area cluster surveys derived from the \rosat \ All Sky Survey \citep{Trumper1993}: the \rosat \ Brightest Cluster Sample \citep{Ebeling1998}; the \rosat-ESO Flux-Limited X-ray Sample \citep{Bohringer2004}; and the MAssive Cluster Survey \citep{Ebeling2007,Ebeling2010}. We also included clusters from the 400-Square Degree {\it ROSAT PSPC} Galaxy Cluster Survey \citep{Burenin2007}. Each sample covers a distinct volume of the Universe: BCS covers the northern sky at $z < 0.3$; REFLEX covers the southern sky at $z < 0.3$; and MACS covers higher redshifts, $0.3 < z <0.9$, at declinations $> -40^{\circ}$. The $400 \deg^{2}$ survey covers high Galactic latitudes at redshifts of $z < 1$. The galaxy clusters included in these samples have been instrumental in recent cosmological studies \citep{Mantz2008,Vikhlinin2009,Mantz2010a,Mantz2010b,Allen2011}. All of the clusters chosen from these samples have \cha \ exposures of at least 10 ks in public archives as well as robust measurements of their masses and virial radii \citep{Mantz2010a,Mantz2010b}, and are a representative sub-sample of these surveys. In total, 135 unique galaxy clusters are included, with redshifts ranging from $0.2 < z < 0.9$. General information for the clusters and the \cha \ data sets used may be found in Table \ref{ChandraSample}. We note that these clusters are among the most massive and X-ray luminous clusters in the Universe, and therefore host large numbers of galaxies and substantial masses of hot, X-ray emitting gas (the Intracluster Medium, hereafter ICM). We therefore expect the influences of the local cluster environment to be pronounced in this sample. With measurements of \rfive \ available for each cluster, we are able to relate observed trends in the AGN population to the virial radii of the clusters.   

\begin{figure}
\includegraphics[width=0.92\columnwidth, angle=270]{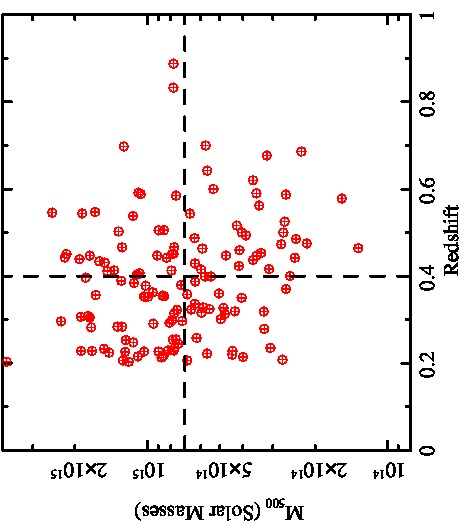}
\caption{\label{MassZDist} The masses and redshifts of the 135 galaxy clusters in this study. The median cluster redshift of $z=0.4$ and cluster mass of $\mysub{M}{500}=7 \times 10^{14} \msolar$ are denoted with dashed lines.   }
\end{figure}

Mass measurements and the associated radii, \rfive, for each cluster are taken from \cite{Mantz2010a,Mantz2010b}.\footnote{The scaling radius \mysub{r}{\Delta} is defined as the radius where the enclosed average mass density is equal to $\Delta$ times the critical density of the universe at the cluster's respective redshift, $\mysub{\rho}{c}(z)$. The corresponding mass \mysub{M}{\Delta} is defined as $\mysub{M}{\Delta} = 4/3 \pi \Delta \mysub{\rho}{c}(z)$ \mysub{r}{\Delta}$^{3}$. The mass range extends from $1 \times 10^{14} \msolar < \mysub{M}{500} < 5 \times 10^{15} \msolar$ and the scaling radii range from $0.6  \Mpc < \mysub{r}{500} < 2 \Mpc$.   \
 } The typical uncertainties in measurements of \rfive \ are of order $ 10\%$. The \rfive \ values and X-ray centroids for the clusters are summarized in Table \ref{ChandraSample}, and the distribution of cluster masses and redshifts used for this study are shown in Figure \ref{MassZDist}.

\begin{table*}
\tiny
\caption{\label{ChandraSample} Summary of the cluster sample and \cha \ observations (either ACIS-I and ACIS-S) listed in order of cluster redshift. A total of 135 clusters of galaxies and 224 independent \cha \ observations were used. The columns list (1) cluster name; (2) redshift; (3) \& (4) right ascension and declination of the cluster X-ray centroid (J2000);  (5) the \cha \ observation number (OBS ID\#); (6) observation date; (7) primary detector array used; (8) exposure time in ks, after cleaning; (9) $\mysub{M}{500}$ for the cluster, in units of $10^{14} \msolar$; (10) \rfive \ in \Mpc; and (11) the equivalent Galactic hydrogen column density in the direction of the cluster, in units of $10^{20} \rm{atoms} \cm^{-2}$. Clusters marked with a $^{\dagger}$ are the clusters included in Paper I.     }
\centering

\begin{tabular}{ c c c c c c c c c c c}\\

  \hline 

(1) & (2) & (3) & (4) & (5) & (6) & (7) & (8) & (9) & (10) & (11) \\
Cluster Name & {\it z}  & RA &  DEC &  OBS ID \# &  Obs Date  &  Detector  & Exposure (ks) & $\mysub{M}{500} (10^{14} \msolar)$ & \rfive (\Mpc) &  \mysub{N}{H} ($10^{20} \cm^{-2}$)  \\ 
\hline\hline

Abell 2163 & 0.203 & $ 16 \ 15 \ 45.96 $ & $ -6 \ 08 \ 53.51 $  & 1653 & 2001-06-16 & ACIS-I & 71.14 &  38.5 & 2.2 & 15.40\\
Abell 2163 & 0.203 & $ 16 \ 15 \ 45.96 $ & $ -6 \ 08 \ 53.51 $  & 545 & 2000-07-29 & ACIS-I & 9.44 &  38.5 & 2.2 & 15.40\\
Abell 520 & 0.203 & $ 04 \ 54 \ 9.02 $ & $ 02 \ 55 \ 17.79 $  & 4215 & 2003-12-04 & ACIS-I & 66.27 &  11.9 & 1.5 & 5.65\\
Abell 520 & 0.203 & $ 04 \ 54 \ 9.02 $ & $ 02 \ 55 \ 17.79 $  & 528 & 2000-10-10 & ACIS-I & 9.46 &  11.9 & 1.5 & 5.65\\
Abell 520 & 0.203 & $ 04 \ 54 \ 9.02 $ & $ 02 \ 55 \ 17.79 $  & 7703 & 2007-01-01 & ACIS-I & 5.08 &  11.9 & 1.5 & 5.65\\
Abell 209$^{\dagger}$ & 0.206 & $ 01 \ 31 \ 53.14 $ & $ -13 \ 36 \ 48.35 $  & 3579 & 2003-08-03 & ACIS-I & 9.98 &  12.6 & 1.5 & 1.44\\
Abell 209$^{\dagger}$ & 0.206 & $ 01 \ 31 \ 53.14 $ & $ -13 \ 36 \ 48.35 $  & 522 & 2000-09-09 & ACIS-I & 9.96 &  12.6 & 1.5 & 1.44\\
Abell 963$^{\dagger}$ & 0.206 & $ 10 \ 17 \ 3.56 $ & $ 39 \ 02 \ 51.51 $  & 7704 & 2007-02-18 & ACIS-I & 5.06 &  6.8 & 1.2 & 1.25\\
Abell 963$^{\dagger}$ & 0.206 & $ 10 \ 17 \ 3.56 $ & $ 39 \ 02 \ 51.51 $  & 903 & 2000-10-11 & ACIS-S & 36.28 &  6.8 & 1.2 & 1.25\\
RX J0439.0+0520 & 0.208 & $ 04 \ 39 \ 2.18 $ & $ 05 \ 20 \ 42.30 $  & 527 & 2000-08-29 & ACIS-I & 9.59 &  2.7 & 0.9 & 8.92\\
RX J0439.0+0520 & 0.208 & $ 04 \ 39 \ 2.18 $ & $ 05 \ 20 \ 42.30 $  & 9369 & 2007-11-12 & ACIS-I & 19.86 &  2.7 & 0.9 & 8.92\\
RX J0439.0+0520 & 0.208 & $ 04 \ 39 \ 2.18 $ & $ 05 \ 20 \ 42.30 $  & 9761 & 2007-11-15 & ACIS-I & 8.65 &  2.7 & 0.9 & 8.92\\
Abell 1423 & 0.213 & $ 11 \ 57 \ 17.35 $ & $ 33 \ 36 \ 39.75 $  & 538 & 2000-07-07 & ACIS-I & 9.87 &  8.7 & 1.3 & 1.81\\
Zwicky 2701 & 0.214 & $ 09 \ 52 \ 49.18 $ & $ 51 \ 53 \ 5.58 $  & 3195 & 2001-11-04 & ACIS-S & 26.91 &  4.0 & 1.0 & 0.75\\
RX J1504.1-0248 & 0.215 & $ 15 \ 04 \ 7.58 $ & $ -2 \ 48 \ 16.30 $  & 4935 & 2004-01-07 & ACIS-I & 13.29 &  11.0 & 1.5 & 5.97\\
RX J1504.1-0248 & 0.215 & $ 15 \ 04 \ 7.58 $ & $ -2 \ 48 \ 16.30 $  & 5793 & 2005-03-20 & ACIS-I & 39.15 &  11.0 & 1.5 & 5.97\\
Abell 773 & 0.217 & $ 09 \ 17 \ 52.68 $ & $ 51 \ 43 \ 35.72 $  & 3588 & 2003-01-25 & ACIS-I & 9.39 &  8.6 & 1.3 & 1.28\\
Abell 773 & 0.217 & $ 09 \ 17 \ 52.68 $ & $ 51 \ 43 \ 35.72 $  & 5006 & 2004-01-21 & ACIS-I & 19.81 &  8.6 & 1.3 & 1.28\\
Abell 773 & 0.217 & $ 09 \ 17 \ 52.68 $ & $ 51 \ 43 \ 35.72 $  & 533 & 2000-09-05 & ACIS-I & 11.25 &  8.6 & 1.3 & 1.28\\
RX J0304.1-3656 & 0.219 & $ 03 \ 04 \ 3.26 $ & $ -36 \ 56 \ 29.54 $  & 9413 & 2008-03-16 & ACIS-I & 19.86 &  4.4 & 1.1 & 2.30\\
RX J0237.4-2630 & 0.222 & $ 02 \ 37 \ 27.42 $ & $ -26 \ 30 \ 27.85 $  & 9412 & 2008-03-03 & ACIS-I & 18.38 &  5.6 & 1.2 & 1.82\\
Abell 2261$^{\dagger}$ & 0.224 & $ 17 \ 22 \ 26.99 $ & $ 32 \ 07 \ 57.89 $  & 5007 & 2004-01-14 & ACIS-I & 24.31 &  14.4 & 1.6 & 3.19\\
Abell 2261$^{\dagger}$ & 0.224 & $ 17 \ 22 \ 26.99 $ & $ 32 \ 07 \ 57.89 $  & 550 & 1999-12-11 & ACIS-I & 9.06 &  14.4 & 1.6 & 3.19\\
Abell 1682 & 0.226 & $ 13 \ 06 \ 50.75 $ & $ 46 \ 33 \ 30.25 $  & 3244 & 2002-10-19 & ACIS-I & 9.77 &  12.4 & 1.5 & 1.04\\
Abell 2667 & 0.226 & $ 23 \ 51 \ 39.72 $ & $ -26 \ 04 \ 59.79 $  & 2214 & 2001-06-19 & ACIS-S & 9.64 &  9.0 & 1.4 & 1.73\\
RX J0638.7-5358 & 0.227 & $ 06 \ 38 \ 47.29 $ & $ -53 \ 58 \ 27.67 $  & 9420 & 2008-04-11 & ACIS-I & 19.89 &  10.3 & 1.4 & 6.06\\
Abell 1763 & 0.228 & $ 13 \ 35 \ 19.03 $ & $ 40 \ 59 \ 59.40 $  & 3591 & 2003-08-28 & ACIS-I & 19.59 &  17.0 & 1.7 & 0.82\\
RX J0220.9-3829 & 0.228 & $ 02 \ 20 \ 56.54 $ & $ -38 \ 28 \ 51.88 $  & 9411 & 2008-02-29 & ACIS-I & 19.86 &  4.4 & 1.1 & 1.96\\
Abell 2219$^{\dagger}$ & 0.228 & $ 16 \ 40 \ 20.35 $ & $ 46 \ 42 \ 30.00 $  & 896 & 2000-03-31 & ACIS-S & 42.29 &  18.9 & 1.7 & 1.76\\
Abell 2111 & 0.229 & $ 15 \ 39 \ 41.13 $ & $ 34 \ 25 \ 7.27 $  & 544 & 2000-03-22 & ACIS-I & 10.29 &  7.8 & 1.3 & 1.84\\
Z5247 & 0.229 & $ 12 \ 34 \ 22.11 $ & $ 09 \ 47 \ 4.93 $  & 539 & 2000-03-23 & ACIS-I & 9.28 &  8.2 & 1.3 & 1.61\\
Abell 2390$^{\dagger}$ & 0.233 & $ 21 \ 53 \ 37.08 $ & $ 17 \ 41 \ 45.39 $  & 4193 & 2003-09-11 & ACIS-S & 95.06 &  15.2 & 1.6 & 6.21\\
Abell 2390$^{\dagger}$ & 0.233 & $ 21 \ 53 \ 37.08 $ & $ 17 \ 41 \ 45.39 $  & 500 & 2000-10-08 & ACIS-S & 9.82 &  15.2 & 1.6 & 6.21\\
Abell 2390$^{\dagger}$ & 0.233 & $ 21 \ 53 \ 37.08 $ & $ 17 \ 41 \ 45.39 $  & 501 & 1999-11-05 & ACIS-S & 9.04 &  15.2 & 1.6 & 6.21\\
Z2089 & 0.235 & $ 09 \ 00 \ 36.84 $ & $ 20 \ 53 \ 40.36 $  & 7897 & 2006-12-23 & ACIS-I & 9.04 &  3.1 & 0.9 & 2.86\\
RX J2129.6+0005$^{\dagger}$ & 0.235 & $ 21 \ 29 \ 39.73 $ & $ 00 \ 05 \ 18.15 $  & 552 & 2000-10-21 & ACIS-I & 9.96 &  7.7 & 1.3 & 3.63\\
RX J2129.6+0005$^{\dagger}$ & 0.235 & $ 21 \ 29 \ 39.73 $ & $ 00 \ 05 \ 18.15 $  & 9370 & 2009-04-03 & ACIS-I & 29.64 &  7.7 & 1.3 & 3.63\\
RX J0439.0+0715 & 0.244 & $ 04 \ 39 \ 0.55 $ & $ 07 \ 16 \ 0.30 $  & 1449 & 1999-10-16 & ACIS-I & 6.31 &  7.4 & 1.3 & 9.18\\
RX J0439.0+0715 & 0.244 & $ 04 \ 39 \ 0.55 $ & $ 07 \ 16 \ 0.30 $  & 3583 & 2003-01-04 & ACIS-I & 19.21 &  7.4 & 1.3 & 9.18\\
RX J0439.0+0715 & 0.244 & $ 04 \ 39 \ 0.55 $ & $ 07 \ 16 \ 0.30 $  & 526 & 1999-10-16 & ACIS-I & 1.59 &  7.4 & 1.3 & 9.18\\
Abell 521$^{\dagger}$ & 0.248 & $ 04 \ 54 \ 7.42 $ & $ -10 \ 13 \ 24.29 $  & 430 & 2000-10-13 & ACIS-S & 39.11 &  11.4 & 1.5 & 4.78\\
Abell 521$^{\dagger}$ & 0.248 & $ 04 \ 54 \ 7.42 $ & $ -10 \ 13 \ 24.29 $  & 901 & 1999-12-23 & ACIS-I & 38.63 &  11.4 & 1.5 & 4.78\\
Abell 1835$^{\dagger}$ & 0.253 & $ 14 \ 01 \ 1.93 $ & $ 02 \ 52 \ 39.89 $  & 6880 & 2006-08-25 & ACIS-I & 117.91 &  12.3 & 1.5 & 2.04\\
Abell 1835$^{\dagger}$ & 0.253 & $ 14 \ 01 \ 1.93 $ & $ 02 \ 52 \ 39.89 $  & 6881 & 2005-12-07 & ACIS-I & 36.28 &  12.3 & 1.5 & 2.04\\
Abell 1835$^{\dagger}$ & 0.253 & $ 14 \ 01 \ 1.93 $ & $ 02 \ 52 \ 39.89 $  & 7370 & 2006-07-24 & ACIS-I & 39.50 &  12.3 & 1.5 & 2.04\\
RX J0307.0-2840 & 0.254 & $ 03 \ 07 \ 1.98 $ & $ -28 \ 39 \ 55.69 $  & 9414 & 2008-03-13 & ACIS-I & 18.90 &  7.8 & 1.3 & 1.27\\
Abell 68$^{\dagger}$ & 0.255 & $ 00 \ 37 \ 5.95 $ & $ 09 \ 09 \ 36.03 $  & 3250 & 2002-09-07 & ACIS-I & 9.98 &  7.6 & 1.3 & 4.96\\
MS1455.0+2232 & 0.258 & $ 14 \ 57 \ 15.07 $ & $ 22 \ 20 \ 33.26 $  & 4192 & 2003-09-05 & ACIS-I & 91.88 &  6.2 & 1.2 & 3.18\\
MS1455.0+2232 & 0.258 & $ 14 \ 57 \ 15.07 $ & $ 22 \ 20 \ 33.26 $  & 543 & 2000-05-19 & ACIS-I & 9.85 &  6.2 & 1.2 & 3.18\\
MS1455.0+2232 & 0.258 & $ 14 \ 57 \ 15.07 $ & $ 22 \ 20 \ 33.26 $  & 7709 & 2007-03-23 & ACIS-I & 7.06 &  6.2 & 1.2 & 3.18\\
RX J2011.3-5725 & 0.279 & $ 20 \ 11 \ 27.24 $ & $ -57 \ 25 \ 10.16 $  & 4995 & 2004-06-08 & ACIS-I & 23.99 &  3.3 & 0.9 & 4.15\\
Abell 697$^{\dagger}$ & 0.282 & $ 08 \ 42 \ 57.62 $ & $ 36 \ 21 \ 57.43 $  & 4217 & 2002-12-15 & ACIS-I & 19.51 &  17.1 & 1.6 & 2.93\\
RX J0232.2-4420 & 0.284 & $ 02 \ 32 \ 17.73 $ & $ -44 \ 20 \ 55.19 $  & 4993 & 2004-06-08 & ACIS-I & 23.40 &  12.7 & 1.5 & 1.69\\
RX J0528.9-3927 & 0.284 & $ 05 \ 28 \ 53.25 $ & $ -39 \ 28 \ 19.12 $  & 4994 & 2004-03-10 & ACIS-I & 22.45 &  13.3 & 1.5 & 2.10\\
Zwicky 3146 & 0.291 & $ 10 \ 23 \ 39.64 $ & $ 04 \ 11 \ 12.17 $  & 909 & 2000-05-10 & ACIS-I & 46.00 &  9.4 & 1.3 & 2.46\\
Zwicky 3146 & 0.291 & $ 10 \ 23 \ 39.64 $ & $ 04 \ 11 \ 12.17 $  & 9371 & 2008-01-18 & ACIS-I & 40.16 &  9.4 & 1.3 & 2.46\\
RX J0043.4-2037 & 0.292 & $ 00 \ 43 \ 24.82 $ & $ -20 \ 37 \ 24.41 $  & 9409 & 2008-02-02 & ACIS-I & 19.91 &  8.1 & 1.3 & 1.84\\
1E0657-56 & 0.296 & $ 06 \ 58 \ 27.55 $ & $ -55 \ 56 \ 32.44 $  & 3184 & 2002-07-12 & ACIS-I & 87.47 &  22.8 & 1.8 & 4.89\\
1E0657-56 & 0.296 & $ 06 \ 58 \ 27.55 $ & $ -55 \ 56 \ 32.44 $  & 4984 & 2004-08-19 & ACIS-I & 76.11 &  22.8 & 1.8 & 4.89\\
1E0657-56 & 0.296 & $ 06 \ 58 \ 27.55 $ & $ -55 \ 56 \ 32.44 $  & 4985 & 2004-08-23 & ACIS-I & 27.48 &  22.8 & 1.8 & 4.89\\
1E0657-56 & 0.296 & $ 06 \ 58 \ 27.55 $ & $ -55 \ 56 \ 32.44 $  & 4986 & 2004-08-25 & ACIS-I & 41.48 &  22.8 & 1.8 & 4.89\\
1E0657-56 & 0.296 & $ 06 \ 58 \ 27.55 $ & $ -55 \ 56 \ 32.44 $  & 5356 & 2004-08-11 & ACIS-I & 97.19 &  22.8 & 1.8 & 4.89\\
1E0657-56 & 0.296 & $ 06 \ 58 \ 27.55 $ & $ -55 \ 56 \ 32.44 $  & 5357 & 2004-08-14 & ACIS-I & 79.05 &  22.8 & 1.8 & 4.89\\
1E0657-56 & 0.296 & $ 06 \ 58 \ 27.55 $ & $ -55 \ 56 \ 32.44 $  & 5358 & 2004-08-15 & ACIS-I & 31.95 &  22.8 & 1.8 & 4.89\\
1E0657-56 & 0.296 & $ 06 \ 58 \ 27.55 $ & $ -55 \ 56 \ 32.44 $  & 5361 & 2004-08-17 & ACIS-I & 82.61 &  22.8 & 1.8 & 4.89\\
1E0657-56 & 0.296 & $ 06 \ 58 \ 27.55 $ & $ -55 \ 56 \ 32.44 $  & 554 & 2000-10-16 & ACIS-I & 25.79 &  22.8 & 1.8 & 4.89\\
Abell 2537$^{\dagger}$ & 0.297 & $ 23 \ 08 \ 22.00 $ & $ -2 \ 11 \ 29.75 $  & 4962 & 2004-09-09 & ACIS-S & 36.19 &  7.2 & 1.2 & 4.62\\
Abell 2537$^{\dagger}$ & 0.297 & $ 23 \ 08 \ 22.00 $ & $ -2 \ 11 \ 29.75 $  & 9372 & 2008-08-11 & ACIS-I & 38.50 &  7.2 & 1.2 & 4.62\\
Abell 781 & 0.298 & $ 09 \ 20 \ 25.34 $ & $ 30 \ 30 \ 10.91 $  & 534 & 2000-10-03 & ACIS-I & 9.94 &  7.9 & 1.3 & 1.65\\
MACS J2245.0+2637 & 0.301 & $ 22 \ 45 \ 4.56 $ & $ 26 \ 38 \ 4.47 $  & 3287 & 2002-11-24 & ACIS-I & 16.86 &  4.9 & 1.1 & 5.04\\

&&&&&&&& \\
\hline

\end{tabular}
\end{table*}

\addtocounter{table}{-1}

\begin{table*}

\tiny
\caption{Continued  }
\centering

\begin{tabular}{ c c c c c c c c c c c}\\

  \hline 

(1) & (2) & (3) & (4) & (5) & (6) & (7) & (8) & (9) & (10) & (11) \\
Cluster Name & {\it z}  & RA &  DEC &  OBS ID \# &  Obs Date  &  Detector  & Exposure (ks) & $\mysub{M}{500} (10^{14} \msolar)$ & \rfive (\Mpc) &  \mysub{N}{H} ($10^{20} \cm^{-2}$)  \\ 
\hline\hline

MACS J2311.5+0338 & 0.305 & $ 23 \ 11 \ 33.17 $ & $ 03 \ 38 \ 7.05 $  & 3288 & 2002-09-07 & ACIS-I & 13.59 &  17.4 & 1.6 & 4.60\\
MACS J1131.8-1955 & 0.306 & $ 11 \ 31 \ 55.61 $ & $ -19 \ 55 \ 45.39 $  & 3276 & 2002-06-14 & ACIS-I & 13.90 &  18.9 & 1.7 & 4.02\\
Abell 2744 & 0.308 & $ 00 \ 14 \ 18.75 $ & $ -30 \ 23 \ 18.04 $  & 2212 & 2001-09-03 & ACIS-S & 24.81 &  17.6 & 1.6 & 1.39\\
Abell 2744 & 0.308 & $ 00 \ 14 \ 18.75 $ & $ -30 \ 23 \ 18.04 $  & 7915 & 2006-11-08 & ACIS-I & 18.61 &  17.6 & 1.6 & 1.39\\
Abell 2744 & 0.308 & $ 00 \ 14 \ 18.75 $ & $ -30 \ 23 \ 18.04 $  & 8477 & 2007-06-10 & ACIS-I & 45.90 &  17.6 & 1.6 & 1.39\\
Abell 2744 & 0.308 & $ 00 \ 14 \ 18.75 $ & $ -30 \ 23 \ 18.04 $  & 8557 & 2007-06-14 & ACIS-I & 27.81 &  17.6 & 1.6 & 1.39\\
MS2137.3-2353$^{\dagger}$ & 0.313 & $ 21 \ 40 \ 15.17 $ & $ -23 \ 39 \ 39.77 $  & 4974 & 2003-11-13 & ACIS-S & 57.38 &  4.7 & 1.1 & 3.76\\
MS2137.3-2353$^{\dagger}$ & 0.313 & $ 21 \ 40 \ 15.17 $ & $ -23 \ 39 \ 39.77 $  & 5250 & 2003-11-18 & ACIS-S & 40.54 &  4.7 & 1.1 & 3.76\\
MS2137.3-2353$^{\dagger}$ & 0.313 & $ 21 \ 40 \ 15.17 $ & $ -23 \ 39 \ 39.77 $  & 928 & 1999-11-18 & ACIS-S & 43.60 &  4.7 & 1.1 & 3.76\\
MACS J0242.5-2132 & 0.314 & $ 02 \ 42 \ 35.88 $ & $ -21 \ 32 \ 26.09 $  & 3266 & 2002-02-07 & ACIS-I & 11.85 &  7.7 & 1.2 & 2.72\\
Abell 1995 & 0.316 & $ 14 \ 52 \ 57.96 $ & $ 58 \ 02 \ 56.33 $  & 7021 & 2006-08-30 & ACIS-I & 48.53 &  5.9 & 1.1 & 1.19\\
Abell 1995 & 0.316 & $ 14 \ 52 \ 57.96 $ & $ 58 \ 02 \ 56.33 $  & 906 & 2000-05-08 & ACIS-S & 45.56 &  5.9 & 1.1 & 1.19\\
MACS J1427.6-2521 & 0.318 & $ 14 \ 27 \ 39.50 $ & $ -25 \ 21 \ 3.05 $  & 3279 & 2002-06-29 & ACIS-I & 16.92 &  3.3 & 0.9 & 5.88\\
MACS J1427.6-2521 & 0.318 & $ 14 \ 27 \ 39.50 $ & $ -25 \ 21 \ 3.05 $  & 9373 & 2008-06-11 & ACIS-I & 28.38 &  3.3 & 0.9 & 5.88\\
MACS J0547.0-3904 & 0.319 & $ 05 \ 47 \ 1.46 $ & $ -39 \ 04 \ 26.05 $  & 3273 & 2002-10-20 & ACIS-I & 21.74 &  4.3 & 1.0 & 3.70\\
MACS J0257.6-2209 & 0.322 & $ 02 \ 57 \ 41.28 $ & $ -22 \ 09 \ 13.42 $  & 3267 & 2001-11-12 & ACIS-I & 20.46 &  7.5 & 1.2 & 2.07\\
MACS J2049.9-3217 & 0.323 & $ 20 \ 49 \ 55.34 $ & $ -32 \ 16 \ 49.39 $  & 3283 & 2002-12-08 & ACIS-I & 23.79 &  6.6 & 1.2 & 5.59\\
MACS J2229.7-2755 & 0.324 & $ 22 \ 29 \ 45.22 $ & $ -27 \ 55 \ 35.98 $  & 3286 & 2002-11-13 & ACIS-I & 16.42 &  5.5 & 1.1 & 1.35\\
MACS J2229.7-2755 & 0.324 & $ 22 \ 29 \ 45.22 $ & $ -27 \ 55 \ 35.98 $  & 9374 & 2007-12-09 & ACIS-I & 14.82 &  5.5 & 1.1 & 1.35\\
MACS J1319.9+7003 & 0.327 & $ 13 \ 20 \ 7.53 $ & $ 70 \ 04 \ 37.21 $  & 3278 & 2002-09-15 & ACIS-I & 21.62 &  4.8 & 1.1 & 1.24\\
Zwicky J1358+6245 & 0.328 & $ 13 \ 59 \ 50.90 $ & $ 62 \ 31 \ 2.89 $  & 516 & 2000-09-03 & ACIS-S & 54.06 &  5.9 & 1.1 & 1.78\\
MACS J0520.7-1328 & 0.336 & $ 05 \ 20 \ 42.17 $ & $ -13 \ 28 \ 46.78 $  & 3272 & 2002-02-10 & ACIS-I & 19.23 &  6.3 & 1.2 & 7.29\\
CL J0302-0423 & 0.350 & $ 03 \ 02 \ 21.06 $ & $ -4 \ 23 \ 23.51 $  & 5782 & 2005-12-07 & ACIS-I & 10.04 &  4.0 & 1.0 & 6.03\\
MACS J1931.8-2634$^{\dagger}$ & 0.352 & $ 19 \ 31 \ 49.61 $ & $ -26 \ 34 \ 33.60 $  & 3282 & 2002-10-20 & ACIS-I & 13.59 &  9.9 & 1.3 & 8.31\\
MACS J1931.8-2634$^{\dagger}$ & 0.352 & $ 19 \ 31 \ 49.61 $ & $ -26 \ 34 \ 33.60 $  & 9382 & 2008-08-21 & ACIS-I & 98.92 &  9.9 & 1.3 & 8.31\\
MACS J0035.4-2015 & 0.352 & $ 00 \ 35 \ 26.22 $ & $ -20 \ 15 \ 46.12 $  & 3262 & 2003-01-22 & ACIS-I & 21.35 &  10.2 & 1.4 & 1.64\\
CL J1212+2733 & 0.353 & $ 12 \ 12 \ 18.40 $ & $ 27 \ 33 \ 1.57 $  & 5767 & 2005-03-17 & ACIS-I & 14.58 &  10.3 & 1.4 & 1.72\\
RBS797 & 0.354 & $ 09 \ 47 \ 13.03 $ & $ 76 \ 23 \ 13.93 $  & 2202 & 2000-10-20 & ACIS-I & 11.74 &  8.5 & 1.3 & 2.28\\
MACS J1115.8+0129$^{\dagger}$ & 0.355 & $ 11 \ 15 \ 51.88 $ & $ 01 \ 29 \ 54.98 $  & 3275 & 2003-01-23 & ACIS-I & 15.90 &  8.6 & 1.3 & 4.34\\
MACS J1115.8+0129$^{\dagger}$ & 0.355 & $ 11 \ 15 \ 51.88 $ & $ 01 \ 29 \ 54.98 $  & 9375 & 2008-02-03 & ACIS-I & 39.62 &  8.6 & 1.3 & 4.34\\
MACS J0308.9+2645 & 0.356 & $ 03 \ 08 \ 56.03 $ & $ 26 \ 45 \ 34.85 $  & 3268 & 2002-03-10 & ACIS-I & 24.44 &  16.4 & 1.6 & 9.43\\
MACS J0404.6+1109 & 0.358 & $ 04 \ 04 \ 32.73 $ & $ 11 \ 08 \ 10.68 $  & 3269 & 2002-02-20 & ACIS-I & 21.81 &  6.8 & 1.2 & 12.30\\
RX J0027.6+2616 & 0.360 & $ 00 \ 27 \ 45.55 $ & $ 26 \ 16 \ 21.75 $  & 3249 & 2002-06-26 & ACIS-I & 9.97 &  5.0 & 1.1 & 3.58\\
RX J1532.9+3021$^{\dagger}$ & 0.363 & $ 15 \ 32 \ 53.83 $ & $ 30 \ 20 \ 59.38 $  & 1649 & 2001-08-26 & ACIS-S & 9.36 &  9.5 & 1.3 & 2.30\\
RX J1532.9+3021$^{\dagger}$ & 0.363 & $ 15 \ 32 \ 53.83 $ & $ 30 \ 20 \ 59.38 $  & 1665 & 2001-09-06 & ACIS-I & 9.97 &  9.5 & 1.3 & 2.30\\
CL J0318-0302 & 0.370 & $ 03 \ 18 \ 33.27 $ & $ -3 \ 02 \ 58.36 $  & 5775 & 2005-03-15 & ACIS-I & 14.57 &  2.6 & 0.8 & 5.30\\
Zwicky J1953$^{\dagger}$ & 0.378 & $ 08 \ 50 \ 6.98 $ & $ 36 \ 04 \ 20.45 $  & 1659 & 2000-10-22 & ACIS-I & 24.86 &  10.2 & 1.3 & 2.96\\
Zwicky J1953$^{\dagger}$ & 0.378 & $ 08 \ 50 \ 6.98 $ & $ 36 \ 04 \ 20.45 $  & 7716 & 2006-12-20 & ACIS-I & 6.98 &  10.2 & 1.3 & 2.96\\
MACS J0011.7-1523 & 0.379 & $ 00 \ 11 \ 42.83 $ & $ -15 \ 23 \ 21.69 $  & 3261 & 2002-11-20 & ACIS-I & 21.60 &  7.2 & 1.2 & 1.85\\
MACS J0011.7-1523 & 0.379 & $ 00 \ 11 \ 42.83 $ & $ -15 \ 23 \ 21.69 $  & 6105 & 2005-06-28 & ACIS-I & 37.27 &  7.2 & 1.2 & 1.85\\
MACS J0949.8+1708$^{\dagger}$ & 0.384 & $ 09 \ 49 \ 51.79 $ & $ 17 \ 07 \ 8.31 $  & 3274 & 2002-11-06 & ACIS-I & 14.31 &  11.3 & 1.4 & 3.08\\
MACS J1720.2+3536$^{\dagger}$ & 0.387 & $ 17 \ 20 \ 16.67 $ & $ 35 \ 36 \ 23.35 $  & 3280 & 2002-11-03 & ACIS-I & 20.84 &  6.3 & 1.1 & 3.46\\
MACS J1720.2+3536$^{\dagger}$ & 0.387 & $ 17 \ 20 \ 16.67 $ & $ 35 \ 36 \ 23.35 $  & 6107 & 2005-11-22 & ACIS-I & 33.88 &  6.3 & 1.1 & 3.46\\
MACS J1731.6+2252$^{\dagger}$ & 0.389 & $ 17 \ 31 \ 39.19 $ & $ 22 \ 51 \ 49.96 $  & 3281 & 2002-11-03 & ACIS-I & 20.50 &  12.8 & 1.4 & 4.99\\
MACS J2211.7-0349$^{\dagger}$ & 0.396 & $ 22 \ 11 \ 45.91 $ & $ -3 \ 49 \ 41.94 $  & 3284 & 2002-10-08 & ACIS-I & 17.73 &  18.1 & 1.6 & 5.53\\
MACS J0429.6-0253$^{\dagger}$ & 0.399 & $ 04 \ 29 \ 36.00 $ & $ -2 \ 53 \ 5.63 $  & 3271 & 2002-02-07 & ACIS-I & 23.16 &  5.8 & 1.1 & 4.34\\
CL J0809+2811 & 0.399 & $ 08 \ 09 \ 41.04 $ & $ 28 \ 12 \ 1.16 $  & 5774 & 2004-11-30 & ACIS-I & 19.68 &  5.4 & 1.1 & 2.98\\
V1416+4446 & 0.400 & $ 14 \ 16 \ 28.06 $ & $ 44 \ 46 \ 42.96 $  & 541 & 1999-12-02 & ACIS-I & 31.15 &  2.5 & 0.8 & 0.76\\
MACS J1006.9+3200 & 0.403 & $ 10 \ 06 \ 54.69 $ & $ 32 \ 01 \ 30.58 $  & 5819 & 2005-01-24 & ACIS-I & 10.88 &  11.1 & 1.4 & 1.52\\
MACS J0159.8-0849 & 0.406 & $ 01 \ 59 \ 49.37 $ & $ -8 \ 49 \ 59.79 $  & 3265 & 2002-10-02 & ACIS-I & 17.90 &  10.8 & 1.3 & 2.06\\
MACS J0159.8-0849 & 0.406 & $ 01 \ 59 \ 49.37 $ & $ -8 \ 49 \ 59.79 $  & 6106 & 2004-12-04 & ACIS-I & 35.30 &  10.8 & 1.3 & 2.06\\
MACS J2228.5+2036$^{\dagger}$ & 0.411 & $ 22 \ 28 \ 32.78 $ & $ 20 \ 37 \ 14.58 $  & 3285 & 2003-01-22 & ACIS-I & 19.85 &  14.7 & 1.5 & 4.26\\
MACS J0152.5-2852 & 0.413 & $ 01 \ 52 \ 33.91 $ & $ -28 \ 53 \ 33.40 $  & 3264 & 2002-09-17 & ACIS-I & 17.54 &  7.9 & 1.2 & 1.51\\
MACS J0159.0-3412 & 0.413 & $ 01 \ 59 \ 2.06 $ & $ -34 \ 13 \ 6.70 $  & 5818 & 2006-02-19 & ACIS-I & 9.42 &  13.7 & 1.5 & 1.51\\
MACS J1105.7-1014 & 0.415 & $ 11 \ 05 \ 45.87 $ & $ -10 \ 14 \ 35.15 $  & 5817 & 2005-01-03 & ACIS-I & 10.32 &  6.0 & 1.1 & 4.10\\
CL J1003+3253 & 0.416 & $ 10 \ 03 \ 4.51 $ & $ 32 \ 53 \ 37.75 $  & 5776 & 2005-03-11 & ACIS-I & 19.85 &  3.1 & 0.9 & 1.68\\
MACS J2046.0-3430 & 0.423 & $ 20 \ 46 \ 0.58 $ & $ -34 \ 30 \ 17.20 $  & 5816 & 2005-06-28 & ACIS-I & 10.03 &  4.2 & 1.0 & 4.59\\
MACS J2046.0-3430 & 0.423 & $ 20 \ 46 \ 0.58 $ & $ -34 \ 30 \ 17.20 $  & 9377 & 2008-06-27 & ACIS-I & 39.23 &  4.2 & 1.0 & 4.59\\
MACS J0451.9+0006$^{\dagger}$ & 0.429 & $ 04 \ 51 \ 54.67 $ & $ 00 \ 06 \ 18.52 $  & 5815 & 2005-01-08 & ACIS-I & 10.21 &  6.3 & 1.1 & 6.85\\
MACS J0553.4-3342 & 0.431 & $ 05 \ 53 \ 25.56 $ & $ -33 \ 42 \ 36.14 $  & 5813 & 2005-01-08 & ACIS-I & 9.94 &  15.1 & 1.5 & 3.32\\
MACS J0358.8-2955 & 0.434 & $ 03 \ 58 \ 53.38 $ & $ -29 \ 55 \ 44.00 $  & 11719 & 2009-10-18 & ACIS-I & 9.64 &  15.8 & 1.5 & 0.98\\
MACS J1226.8+2153 & 0.437 & $ 12 \ 26 \ 51.04 $ & $ 21 \ 49 \ 54.98 $  & 12878 & 2011-04-11 & ACIS-I & 129.97 &  0.0 & 0.9 & 1.66\\
MACS J1226.8+2153 & 0.437 & $ 12 \ 26 \ 51.04 $ & $ 21 \ 49 \ 54.98 $  & 3590 & 2003-12-13 & ACIS-I & 19.00 &  3.6 & 0.9 & 1.66\\

&&&&&&&& \\
\hline

\end{tabular}
\end{table*}

\addtocounter{table}{-1}

\begin{table*}
\tiny

\caption{Continued  }
\centering

\begin{tabular}{ c c c c c c c c c c c}\\

  \hline 

(1) & (2) & (3) & (4) & (5) & (6) & (7) & (8) & (9) & (10) & (11) \\
Cluster Name & {\it z}  & RA &  DEC &  OBS ID \# &  Obs Date  &  Detector  & Exposure (ks) & $\mysub{M}{500} (10^{14} \msolar)$ & \rfive (\Mpc) &  \mysub{N}{H} ($10^{20} \cm^{-2}$)  \\ 
\hline\hline

MACS J1206.2-0847$^{\dagger}$ & 0.439 & $ 12 \ 06 \ 12.29 $ & $ -8 \ 48 \ 6.22 $  & 3277 & 2002-12-15 & ACIS-I & 23.45 &  19.2 & 1.6 & 4.35\\
CL J0141-3034 & 0.442 & $ 01 \ 41 \ 32.95 $ & $ -30 \ 34 \ 41.72 $  & 5778 & 2005-06-04 & ACIS-I & 29.65 &  2.4 & 0.8 & 1.65\\
IRAS09104 & 0.442 & $ 09 \ 13 \ 45.37 $ & $ 40 \ 56 \ 27.46 $  & 509 & 1999-11-03 & ACIS-S & 9.05 &  8.3 & 1.2 & 1.42\\
MACS J0417.5-1154$^{\dagger}$ & 0.443 & $ 04 \ 17 \ 34.32 $ & $ -11 \ 54 \ 26.65 $  & 11759 & 2009-10-28 & ACIS-I & 51.35 &  22.1 & 1.7 & 3.31\\
MACS J0417.5-1154$^{\dagger}$ & 0.443 & $ 04 \ 17 \ 34.32 $ & $ -11 \ 54 \ 26.65 $  & 12010 & 2009-10-29 & ACIS-I & 25.78 &  22.1 & 1.7 & 3.31\\
MACS J0417.5-1154$^{\dagger}$ & 0.443 & $ 04 \ 17 \ 34.32 $ & $ -11 \ 54 \ 26.65 $  & 3270 & 2002-03-10 & ACIS-I & 12.01 &  22.1 & 1.7 & 3.31\\
MACS J2243.3-0935$^{\dagger}$ & 0.447 & $ 22 \ 43 \ 21.43 $ & $ -9 \ 35 \ 42.76 $  & 3260 & 2002-12-23 & ACIS-I & 20.50 &  17.4 & 1.6 & 4.02\\
MACS J0455.2+0657 & 0.447 & $ 04 \ 55 \ 17.28 $ & $ 06 \ 57 \ 47.60 $  & 5812 & 2005-01-08 & ACIS-I & 9.94 &  9.1 & 1.2 & 8.41\\
MACS J1359.1-1929 & 0.447 & $ 13 \ 59 \ 10.30 $ & $ -19 \ 29 \ 23.36 $  & 5811 & 2005-03-17 & ACIS-I & 9.91 &  3.5 & 0.9 & 5.99\\
MACS J0326.8-0043 & 0.447 & $ 03 \ 26 \ 49.99 $ & $ 00 \ 43 \ 52.20 $  & 5810 & 2005-10-30 & ACIS-I & 9.91 &  4.7 & 1.0 & 6.76\\
MACS J0329.6-0211$^{\dagger}$ & 0.450 & $ 03 \ 29 \ 41.46 $ & $ -2 \ 11 \ 45.52 $  & 3257 & 2001-11-25 & ACIS-I & 9.86 &  7.9 & 1.2 & 4.64\\
MACS J0329.6-0211$^{\dagger}$ & 0.450 & $ 03 \ 29 \ 41.46 $ & $ -2 \ 11 \ 45.52 $  & 3582 & 2002-12-24 & ACIS-I & 19.84 &  7.9 & 1.2 & 4.64\\
MACS J0329.6-0211$^{\dagger}$ & 0.450 & $ 03 \ 29 \ 41.46 $ & $ -2 \ 11 \ 45.52 $  & 6108 & 2004-12-06 & ACIS-I & 39.64 &  7.9 & 1.2 & 4.64\\
MACS J0329.6-0211$^{\dagger}$ & 0.450 & $ 03 \ 29 \ 41.46 $ & $ -2 \ 11 \ 45.52 $  & 7719 & 2006-12-03 & ACIS-I & 7.08 &  7.9 & 1.2 & 4.64\\
RX J1347.5-1145$^{\dagger}$ & 0.451 & $ 13 \ 47 \ 30.77 $ & $ -11 \ 45 \ 9.43 $  & 3592 & 2003-09-03 & ACIS-I & 57.71 &  21.7 & 1.7 & 4.60\\
RX J1347.5-1145$^{\dagger}$ & 0.451 & $ 13 \ 47 \ 30.77 $ & $ -11 \ 45 \ 9.43 $  & 506 & 2000-03-05 & ACIS-S & 8.93 &  21.7 & 1.7 & 4.60\\
RX J1347.5-1145$^{\dagger}$ & 0.451 & $ 13 \ 47 \ 30.77 $ & $ -11 \ 45 \ 9.43 $  & 507 & 2000-04-29 & ACIS-S & 9.99 &  21.7 & 1.7 & 4.60\\
MACS J0140.0-0555 & 0.451 & $ 01 \ 40 \ 1.06 $ & $ -5 \ 55 \ 7.39 $  & 5013 & 2004-06-04 & ACIS-I & 10.19 &  7.8 & 1.2 & 2.75\\
V1701+6414 & 0.453 & $ 17 \ 01 \ 23.10 $ & $ 64 \ 14 \ 7.86 $  & 547 & 2000-10-31 & ACIS-I & 49.52 &  3.4 & 0.9 & 2.28\\
3C295 & 0.460 & $ 14 \ 11 \ 20.34 $ & $ 52 \ 12 \ 10.52 $  & 2254 & 2001-05-18 & ACIS-I & 90.95 &  4.1 & 1.0 & 1.34\\
3C295 & 0.460 & $ 14 \ 11 \ 20.34 $ & $ 52 \ 12 \ 10.52 $  & 578 & 1999-08-30 & ACIS-S & 18.79 &  4.1 & 1.0 & 1.34\\
MACS J1621.3+3810$^{\dagger}$ & 0.463 & $ 16 \ 21 \ 24.75 $ & $ 38 \ 10 \ 9.31 $  & 10785 & 2008-10-18 & ACIS-I & 29.75 &  5.9 & 1.1 & 1.13\\
MACS J1621.3+3810$^{\dagger}$ & 0.463 & $ 16 \ 21 \ 24.75 $ & $ 38 \ 10 \ 9.31 $  & 3254 & 2002-10-18 & ACIS-I & 9.84 &  5.9 & 1.1 & 1.13\\
MACS J1621.3+3810$^{\dagger}$ & 0.463 & $ 16 \ 21 \ 24.75 $ & $ 38 \ 10 \ 9.31 $  & 6109 & 2004-12-11 & ACIS-I & 37.54 &  5.9 & 1.1 & 1.13\\
MACS J1621.3+3810$^{\dagger}$ & 0.463 & $ 16 \ 21 \ 24.75 $ & $ 38 \ 10 \ 9.31 $  & 6172 & 2004-12-25 & ACIS-I & 29.75 &  5.9 & 1.1 & 1.13\\
MACS J1621.3+3810$^{\dagger}$ & 0.463 & $ 16 \ 21 \ 24.75 $ & $ 38 \ 10 \ 9.31 $  & 9379 & 2008-10-17 & ACIS-I & 29.91 &  5.9 & 1.1 & 1.13\\
CL J1641+4001 & 0.464 & $ 16 \ 41 \ 53.42 $ & $ 40 \ 01 \ 45.21 $  & 3575 & 2003-09-24 & ACIS-I & 46.52 &  1.3 & 0.7 & 1.04\\
MACS J1115.2+5320 & 0.466 & $ 11 \ 15 \ 15.09 $ & $ 53 \ 19 \ 56.18 $  & 3253 & 2002-03-23 & ACIS-I & 8.77 &  12.7 & 1.4 & 0.89\\
MACS J1115.2+5320 & 0.466 & $ 11 \ 15 \ 15.09 $ & $ 53 \ 19 \ 56.18 $  & 5008 & 2004-06-22 & ACIS-I & 17.98 &  12.7 & 1.4 & 0.89\\
MACS J1115.2+5320 & 0.466 & $ 11 \ 15 \ 15.09 $ & $ 53 \ 19 \ 56.18 $  & 5350 & 2004-07-28 & ACIS-I & 6.87 &  12.7 & 1.4 & 0.89\\
MACS J1108.8+0906$^{\dagger}$ & 0.466 & $ 11 \ 08 \ 55.15 $ & $ 09 \ 06 \ 2.79 $  & 3252 & 2002-11-17 & ACIS-I & 9.94 &  7.7 & 1.2 & 2.22\\
MACS J1108.8+0906$^{\dagger}$ & 0.466 & $ 11 \ 08 \ 55.15 $ & $ 09 \ 06 \ 2.79 $  & 5009 & 2004-02-20 & ACIS-I & 24.46 &  7.7 & 1.2 & 2.22\\
CL J0355-3741 & 0.473 & $ 03 \ 55 \ 59.45 $ & $ -37 \ 41 \ 45.55 $  & 5761 & 2006-01-12 & ACIS-I & 27.68 &  2.8 & 0.8 & 1.19\\
CL J0333-2456 & 0.475 & $ 03 \ 33 \ 10.75 $ & $ -24 \ 56 \ 31.27 $  & 5764 & 2005-04-05 & ACIS-I & 43.59 &  2.2 & 0.8 & 1.24\\
MACS J0111.5+0855 & 0.485 & $ 01 \ 11 \ 31.36 $ & $ 08 \ 55 \ 40.35 $  & 3256 & 2002-11-20 & ACIS-I & 19.38 &  2.4 & 0.8 & 4.52\\
MACS J1427.2+4407$^{\dagger}$ & 0.487 & $ 14 \ 27 \ 16.02 $ & $ 44 \ 07 \ 30.51 $  & 6112 & 2005-02-12 & ACIS-I & 9.38 &  6.3 & 1.1 & 1.19\\
MACS J1427.2+4407$^{\dagger}$ & 0.487 & $ 14 \ 27 \ 16.02 $ & $ 44 \ 07 \ 30.51 $  & 9380 & 2008-01-14 & ACIS-I & 25.81 &  6.3 & 1.1 & 1.19\\
MACS J1427.2+4407$^{\dagger}$ & 0.487 & $ 14 \ 27 \ 16.02 $ & $ 44 \ 07 \ 30.51 $  & 9808 & 2008-01-15 & ACIS-I & 14.93 &  6.3 & 1.1 & 1.19\\
MACS J1311.0-0310 & 0.494 & $ 13 \ 11 \ 1.69 $ & $ -3 \ 10 \ 39.95 $  & 3258 & 2002-12-15 & ACIS-I & 14.91 &  3.9 & 0.9 & 1.82\\
MACS J1311.0-0310 & 0.494 & $ 13 \ 11 \ 1.69 $ & $ -3 \ 10 \ 39.95 $  & 6110 & 2005-04-20 & ACIS-I & 63.20 &  3.9 & 0.9 & 1.82\\
MACS J1311.0-0310 & 0.494 & $ 13 \ 11 \ 1.69 $ & $ -3 \ 10 \ 39.95 $  & 7721 & 2007-03-03 & ACIS-I & 7.05 &  3.9 & 0.9 & 1.82\\
MACS J1311.0-0310 & 0.494 & $ 13 \ 11 \ 1.69 $ & $ -3 \ 10 \ 39.95 $  & 9381 & 2007-12-09 & ACIS-I & 29.73 &  3.9 & 0.9 & 1.82\\
CL J1002+6858 & 0.500 & $ 10 \ 02 \ 8.99 $ & $ 68 \ 58 \ 35.55 $  & 5773 & 2005-01-05 & ACIS-I & 19.79 &  4.0 & 0.9 & 5.19\\
RX J003033.2+261819 & 0.500 & $ 00 \ 30 \ 33.84 $ & $ 26 \ 18 \ 8.69 $  & 5762 & 2005-05-28 & ACIS-I & 17.88 &  2.7 & 0.8 & 3.71\\
MACS J2214.9-1359$^{\dagger}$ & 0.502 & $ 22 \ 14 \ 57.31 $ & $ -14 \ 00 \ 11.39 $  & 3259 & 2002-12-22 & ACIS-I & 19.47 &  13.2 & 1.4 & 2.88\\
MACS J2214.9-1359$^{\dagger}$ & 0.502 & $ 22 \ 14 \ 57.31 $ & $ -14 \ 00 \ 11.39 $  & 5011 & 2003-11-17 & ACIS-I & 18.52 &  13.2 & 1.4 & 2.88\\
MACS J0911.2+1746$^{\dagger}$ & 0.505 & $ 09 \ 11 \ 10.87 $ & $ 17 \ 46 \ 31.38 $  & 3587 & 2003-02-23 & ACIS-I & 17.87 &  9.0 & 1.2 & 3.28\\
MACS J0911.2+1746$^{\dagger}$ & 0.505 & $ 09 \ 11 \ 10.87 $ & $ 17 \ 46 \ 31.38 $  & 5012 & 2004-03-08 & ACIS-I & 23.79 &  9.0 & 1.2 & 3.28\\
MACS J0257.1-2325$^{\dagger}$ & 0.505 & $ 02 \ 57 \ 9.10 $ & $ -23 \ 26 \ 3.90 $  & 1654 & 2000-10-03 & ACIS-I & 19.84 &  8.5 & 1.2 & 2.08\\
MACS J0257.1-2325$^{\dagger}$ & 0.505 & $ 02 \ 57 \ 9.10 $ & $ -23 \ 26 \ 3.90 $  & 3581 & 2003-08-23 & ACIS-I & 18.47 &  8.5 & 1.2 & 2.08\\
V1525+0958 & 0.516 & $ 15 \ 24 \ 39.78 $ & $ 09 \ 57 \ 46.07 $  & 1664 & 2002-04-01 & ACIS-I & 50.87 &  4.2 & 0.9 & 2.72\\
CL J1357+6232 & 0.525 & $ 13 \ 57 \ 17.64 $ & $ 62 \ 32 \ 50.80 $  & 5763 & 2006-01-24 & ACIS-I & 25.68 &  2.7 & 0.8 & 1.83\\
CL J1357+6232 & 0.525 & $ 13 \ 57 \ 17.64 $ & $ 62 \ 32 \ 50.80 $  & 7267 & 2006-01-29 & ACIS-I & 18.21 &  2.7 & 0.8 & 1.83\\
MACS J0454.1-0300$^{\dagger}$ & 0.538 & $ 04 \ 54 \ 11.45 $ & $ -3 \ 00 \ 50.76 $  & 529 & 2000-01-14 & ACIS-I & 13.90 &  11.5 & 1.3 & 3.92\\
MACS J0454.1-0300$^{\dagger}$ & 0.538 & $ 04 \ 54 \ 11.45 $ & $ -3 \ 00 \ 50.76 $  & 902 & 2000-10-08 & ACIS-S & 44.19 &  11.5 & 1.3 & 3.92\\
MACS J1423.8+2404$^{\dagger}$ & 0.543 & $ 14 \ 23 \ 47.92 $ & $ 24 \ 04 \ 42.77 $  & 1657 & 2001-06-01 & ACIS-I & 18.52 &  6.6 & 1.1 & 2.20\\
MACS J1423.8+2404$^{\dagger}$ & 0.543 & $ 14 \ 23 \ 47.92 $ & $ 24 \ 04 \ 42.77 $  & 4195 & 2003-08-18 & ACIS-S & 115.57 &  6.6 & 1.1 & 2.20\\
MACS J1149.5+2223$^{\dagger}$ & 0.544 & $ 11 \ 49 \ 35.42 $ & $ 22 \ 24 \ 3.62 $  & 1656 & 2001-06-01 & ACIS-I & 18.52 &  18.7 & 1.5 & 1.92\\
MACS J1149.5+2223$^{\dagger}$ & 0.544 & $ 11 \ 49 \ 35.42 $ & $ 22 \ 24 \ 3.62 $  & 3589 & 2003-02-07 & ACIS-I & 20.04 &  18.7 & 1.5 & 1.92\\
MACS J0717.5+3745$^{\dagger}$ & 0.546 & $ 07 \ 17 \ 32.09 $ & $ 37 \ 45 \ 20.94 $  & 1655 & 2001-01-29 & ACIS-I & 19.87 &  24.9 & 1.7 & 6.64\\
MACS J0717.5+3745$^{\dagger}$ & 0.546 & $ 07 \ 17 \ 32.09 $ & $ 37 \ 45 \ 20.94 $  & 4200 & 2003-01-08 & ACIS-I & 59.16 &  24.9 & 1.7 & 6.64\\
MS0015.9+1609$^{\dagger}$ & 0.547 & $ 00 \ 18 \ 33.45 $ & $ 16 \ 26 \ 13.00 $  & 520 & 2000-08-18 & ACIS-I & 67.41 &  16.5 & 1.5 & 3.99\\
V1121+2327 & 0.562 & $ 11 \ 20 \ 57.54 $ & $ 23 \ 26 \ 28.93 $  & 1660 & 2001-04-23 & ACIS-I & 71.24 &  3.4 & 0.8 & 1.14\\
CL J0216-1747 & 0.578 & $ 02 \ 16 \ 33.33 $ & $ -17 \ 47 \ 31.88 $  & 5760 & 2005-09-07 & ACIS-I & 40.04 &  1.5 & 0.7 & 3.02\\
CL J0216-1747 & 0.578 & $ 02 \ 16 \ 33.33 $ & $ -17 \ 47 \ 31.88 $  & 6393 & 2005-10-04 & ACIS-I & 26.64 &  1.5 & 0.7 & 3.02\\
MACS J0025.4-1222$^{\dagger}$ & 0.585 & $ 00 \ 25 \ 29.91 $ & $ -12 \ 22 \ 44.64 $  & 10413 & 2008-10-16 & ACIS-I & 75.63 &  7.6 & 1.1 & 2.50\\
MACS J0025.4-1222$^{\dagger}$ & 0.585 & $ 00 \ 25 \ 29.91 $ & $ -12 \ 22 \ 44.64 $  & 10786 & 2008-10-18 & ACIS-I & 14.12 &  7.6 & 1.1 & 2.50\\
MACS J0025.4-1222$^{\dagger}$ & 0.585 & $ 00 \ 25 \ 29.91 $ & $ -12 \ 22 \ 44.64 $  & 10797 & 2008-10-21 & ACIS-I & 23.85 &  7.6 & 1.1 & 2.50\\
MACS J0025.4-1222$^{\dagger}$ & 0.585 & $ 00 \ 25 \ 29.91 $ & $ -12 \ 22 \ 44.64 $  & 3251 & 2002-11-11 & ACIS-I & 19.32 &  7.6 & 1.1 & 2.50\\
MACS J0025.4-1222$^{\dagger}$ & 0.585 & $ 00 \ 25 \ 29.91 $ & $ -12 \ 22 \ 44.64 $  & 5010 & 2004-08-09 & ACIS-I & 24.82 &  7.6 & 1.1 & 2.50\\
CL J0956+4107 & 0.587 & $ 09 \ 56 \ 3.34 $ & $ 41 \ 07 \ 8.08 $  & 5294 & 2003-12-30 & ACIS-I & 17.34 &  2.6 & 0.8 & 1.22\\
CL J0956+4107 & 0.587 & $ 09 \ 56 \ 3.34 $ & $ 41 \ 07 \ 8.08 $  & 5759 & 2005-01-28 & ACIS-I & 40.16 &  2.6 & 0.8 & 1.22\\
MACS J2129.4-0741$^{\dagger}$ & 0.588 & $ 21 \ 29 \ 25.72 $ & $ -7 \ 41 \ 30.84 $  & 3199 & 2002-12-23 & ACIS-I & 19.85 &  10.6 & 1.3 & 4.33\\
MACS J2129.4-0741$^{\dagger}$ & 0.588 & $ 21 \ 29 \ 25.72 $ & $ -7 \ 41 \ 30.84 $  & 3595 & 2003-10-18 & ACIS-I & 19.87 &  10.6 & 1.3 & 4.33\\
CL J0328-2140 & 0.590 & $ 03 \ 28 \ 36.20 $ & $ -21 \ 40 \ 22.36 $  & 5755 & 2005-03-15 & ACIS-I & 43.29 &  3.5 & 0.9 & 2.11\\
CL J0328-2140 & 0.590 & $ 03 \ 28 \ 36.20 $ & $ -21 \ 40 \ 22.36 $  & 6258 & 2005-03-18 & ACIS-I & 13.09 &  3.5 & 0.9 & 2.11\\
MACS J0647.7+7015$^{\dagger}$ & 0.592 & $ 06 \ 47 \ 49.68 $ & $ 70 \ 14 \ 56.05 $  & 3196 & 2002-10-31 & ACIS-I & 19.27 &  10.9 & 1.3 & 5.40\\
MACS J0647.7+7015$^{\dagger}$ & 0.592 & $ 06 \ 47 \ 49.68 $ & $ 70 \ 14 \ 56.05 $  & 3584 & 2003-10-07 & ACIS-I & 19.99 &  10.9 & 1.3 & 5.40\\
CL J1120+4318 & 0.600 & $ 11 \ 20 \ 6.93 $ & $ 43 \ 18 \ 5.01 $  & 5771 & 2005-01-11 & ACIS-I & 19.83 &  5.3 & 1.0 & 2.97\\
CL J1334+5031 & 0.620 & $ 13 \ 34 \ 20.13 $ & $ 50 \ 31 \ 0.93 $  & 5772 & 2005-08-05 & ACIS-I & 19.49 &  3.6 & 0.9 & 1.05\\
CL J0542.8-4100 & 0.642 & $ 05 \ 42 \ 50.11 $ & $ -41 \ 00 \ 3.53 $  & 914 & 2000-07-26 & ACIS-I & 50.40 &  5.6 & 1.0 & 3.18\\
CL J1202+5751 & 0.677 & $ 12 \ 02 \ 17.79 $ & $ 57 \ 51 \ 53.89 $  & 5757 & 2005-09-02 & ACIS-I & 58.98 &  3.2 & 0.8 & 1.74\\
CL J0405-4100 & 0.686 & $ 04 \ 05 \ 24.52 $ & $ -41 \ 00 \ 19.38 $  & 5756 & 2005-10-27 & ACIS-I & 7.94 &  2.3 & 0.7 & 1.27\\
CL J0405-4100 & 0.686 & $ 04 \ 05 \ 24.52 $ & $ -41 \ 00 \ 19.38 $  & 7191 & 2006-05-19 & ACIS-I & 69.21 &  2.3 & 0.7 & 1.27\\
MACS J0744.8+3927$^{\dagger}$ & 0.698 & $ 07 \ 44 \ 52.32 $ & $ 39 \ 27 \ 26.80 $  & 3197 & 2001-11-12 & ACIS-I & 20.23 &  12.5 & 1.3 & 5.66\\
MACS J0744.8+3927$^{\dagger}$ & 0.698 & $ 07 \ 44 \ 52.32 $ & $ 39 \ 27 \ 26.80 $  & 3585 & 2003-01-04 & ACIS-I & 19.85 &  12.5 & 1.3 & 5.66\\
MACS J0744.8+3927$^{\dagger}$ & 0.698 & $ 07 \ 44 \ 52.32 $ & $ 39 \ 27 \ 26.80 $  & 6111 & 2004-12-03 & ACIS-I & 49.50 &  12.5 & 1.3 & 5.66\\
V1221+4918 & 0.700 & $ 12 \ 21 \ 26.14 $ & $ 49 \ 18 \ 30.69 $  & 1662 & 2001-08-05 & ACIS-I & 79.08 &  5.7 & 1.0 & 1.54\\
CL J0152.7-1357 & 0.833 & $ 01 \ 52 \ 41.16 $ & $ -13 \ 58 \ 6.92 $  & 913 & 2000-09-08 & ACIS-I & 36.48 &  7.8 & 1.0 & 1.33\\
CL J1226.9+3332 & 0.888 & $ 12 \ 26 \ 57.91 $ & $ 33 \ 32 \ 48.60 $  & 3180 & 2003-01-27 & ACIS-I & 31.69 &  7.8 & 1.0 & 1.83\\
CL J1226.9+3332 & 0.888 & $ 12 \ 26 \ 57.91 $ & $ 33 \ 32 \ 48.60 $  & 5014 & 2004-08-07 & ACIS-I & 32.71 &  7.8 & 1.0 & 1.83\\
CL J1226.9+3332 & 0.888 & $ 12 \ 26 \ 57.91 $ & $ 33 \ 32 \ 48.60 $  & 932 & 2000-07-31 & ACIS-S & 9.82 &  7.8 & 1.0 & 1.83\\

&&&&&&&& \\
\hline

\end{tabular}
\end{table*}

\addtocounter{table}{-1}

\section{\cha \ Data Reduction and Catalog Production}\label{ImageEmap}

All of the galaxy cluster fields were observed with the Advanced CCD Imaging Spectrometer (ACIS) aboard \cha. The data were analyzed using the same analysis procedure discussed in Paper I. This includes the initial processing of each event list, image creation, candidate source detection, and refinement of the source catalogs using the \ae \ point-source analysis package\footnote{
  The {\em ACIS Extract} software package and User's Guide are available at
  http://www.astro.psu.edu/xray/acis/acis\_analysis.html. 
  } 
\citep{Broos2010}. Our \ae \ analysis pipeline has three main stages and follows the study of the 4-Ms \cdfs \ \citep{Xue2011}, modified to accommodate the higher background rates and shorter exposure times of our cluster observations. Tests discussed in Paper I have shown that our analysis efficiently rejects spurious detections while preserving high completeness. We perform a similar analysis in each of three energy bands: a soft band ($0.5-2.0 \keV$), a hard band ($2.0-8.0 \keV$), and the full band ($0.5-8.0 \keV$). 

Every cluster field was subsequently inspected visually to ensure that candidate sources associated with cluster substructure were removed from the analysis. Since cold fronts, cool cores, and cavities associated with mechanical feedback from AGN are all sources of potential surface brightness fluctuations on spatial scales comparable to \cha \ PSF, these regions can only be rejected reliably by such visual inspection. We adopt the same criterion for inclusion in the final catalog as in Paper I, including all sources that satisfy a no-source binomial probability threshold of \thresh \ in any of the three energy bands. A total of 11671 sources satisfy this threshold. 

11328, 9244, and 7128 sources satisfy \thresh \ in the full band, soft band, and hard band, respectively, and 5448 sources satisfy this same threshold in all three bands. The majority of these sources are detected in more than one energy band. Only 747, 247, and 96 sources are detected exclusively in the full band, soft band, and hard band respectively.

\subsection{Sensitivity Maps}

Determinations of the local flux limit to which we can robustly identify a point source, commonly known as the sensitivity map, were performed in an identical manner to that discussed in Paper I. In short, our procedure solves for the number of counts required to satisfy our no-source binomial probability threshold, \thresh, for each position in the field of view. Our procedure takes into account PSF broadening with off-axis angle, variations in the effective exposure time (due to vignetting and CCD chip gaps, for example), and variations in the background across the field of view, including the diffuse galaxy cluster emission. 

\begin{figure}
\includegraphics[width=0.92\columnwidth, angle=270]{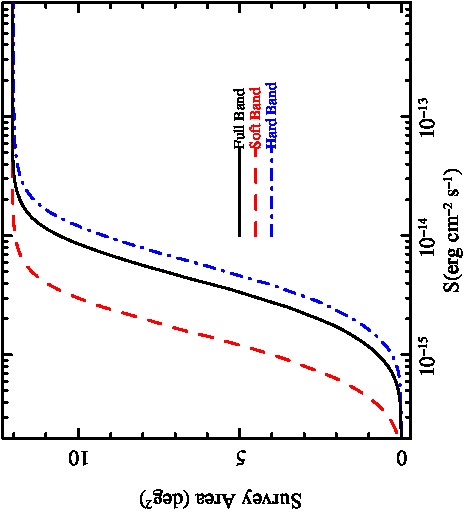}
\caption{\label{SurveyArea} Survey solid angle as a function of flux limit in the soft, hard, and full bands for all cluster observations. We only include the central 12 arcminutes of each pointing in calculating the $\log{N}-\log{S}$ and radial profiles, in order to ensure sample completeness. The total survey area over these 135 clusters is $12.0 \deg^{2}$.  }
\end{figure}

Only a small number of sources in the final catalog have measured full-band fluxes below the flux limit at their respective positions ($371/10246 \sim 3.6\%$). In the majority of these cases, the flux measurements are consistent with the flux limits within statistical uncertainties. Only 128 (1.2\%) of the point sources in the final catalog have a measured flux inconsistent with the local flux limit at their respective positions at a level greater than $68\%$ confidence.\footnote{Uncertainties on the source fluxes are estimated at the $\sim 30\%$ level. This is the typical uncertainty in the flux measured from spectral fits to these sources using {\small XSPEC}. } It can be expected that some sources will have such characteristics, given differences in the spectra of the point sources relative to the assumed AGN spectrum (a power-law with photon index of $\Gamma=1.4$). Indeed, those sources that are fainter than their local flux limits for which successful spectral fits were obtained are measured to be significantly softer than our canonical AGN source. General information about the point source catalogs and sensitivity maps for each cluster can be found in Table \ref{ChandraProperties}. The total survey area sensitive to different flux levels is shown in Figure \ref{SurveyArea}. In order to ensure that our sample is reliably complete, we restrict ourselves to the central 12 arcminutes of each field of view.

\begin{table}
\small
\caption{\label{ChandraProperties}  Summary of X-ray point source numbers and flux limits for each cluster field. The columns list (1) cluster name; (2) the number of AGN satisfying \thresh \ in the soft band, the hard band, the full band, and in any band, respectively; (3) the flux limit for each cluster observation in the soft, hard, and full bands, defined as the minimum flux to which 50\% of the survey area is sensitive, in units of $\times 10^{-15} \ergpcmsqps$. Observations denoted with an $^{a}$ utilize a mix of ACIS-S and ACIS-I observations.  }
\centering
\begin{tabular*}{0.9\columnwidth}{ccc}\\
  \hline Cluster Name  & \mysub{N}{soft}/\mysub{N}{hard}/\mysub{N}{full}/\mysub{N}{any} &  Flux Limit  \\ 
\hline\hline
Abell 2163 & 68/57/83/90 & 1.12/3.55/2.82\\
Abell 520 & 95/78/121/133 & 1.26/5.62/3.98\\
Abell 209 & 36/33/50/56 & 2.00/6.31/5.01\\
Abell 963$^{a}$ & 61/45/75/79 & 2.51/14.13/8.91\\
RX J0439.0+0520 & 76/72/103/107 & 1.26/5.01/3.98\\
Abell 1423 & 27/18/33/39 & 2.82/10.00/7.08\\
Zwicky 2701 & 49/29/58/62 & 1.26/5.01/3.55\\
RX J1504.1-0248 & 73/54/88/93 & 1.12/4.47/3.16\\
Abell 773 & 72/72/96/103 & 1.78/7.08/5.01\\
RX J0304.1-3656 & 40/30/55/59 & 1.78/7.08/5.01\\
RX J0237.4-2630 & 43/31/53/54 & 2.00/7.94/5.62\\
Abell 2261 & 59/48/68/75 & 1.58/5.62/3.98\\
Abell 1682 & 33/24/40/51 & 3.55/12.59/10.00\\
Abell 2667 & 23/19/29/34 & 3.16/11.22/7.94\\
RX J0638.7-5358 & 44/36/54/60 & 2.00/7.08/5.01\\
Abell 1763 & 47/38/59/64 & 2.00/6.31/5.01\\
RX J0220.9-3829 & 52/41/63/65 & 1.78/7.08/5.01\\
Abell 2219 & 31/18/34/40 & 1.26/4.47/3.16\\
Abell 2111 & 38/19/46/50 & 3.16/10.00/7.94\\
Z5247 & 30/21/36/36 & 3.55/11.22/7.94\\
Abell 2390 & 69/54/88/93 & 1.58/10.00/5.62\\
Z2089 & 29/19/37/37 & 2.82/11.22/7.94\\
RX J2129.6+0005 & 81/54/99/109 & 1.78/7.08/5.01\\
RX J0439.0+0715 & 59/42/69/86 & 2.51/10.00/7.08\\
Abell 521$^{a}$ & 110/81/133/141 & 1.00/4.47/3.16\\
Abell 1835 & 137/115/171/177 & 0.79/3.16/2.51\\
RX J0307.0-2840 & 55/29/61/65 & 2.00/7.08/5.01\\
Abell 68 & 36/30/47/50 & 2.82/10.00/7.08\\
MS1455.0+2232 & 130/96/142/158 & 0.89/3.55/2.51\\
RX J2011.3-5725 & 65/57/79/82 & 1.41/5.62/3.98\\
Abell 697 & 51/42/66/69 & 1.78/6.31/5.01\\
RX J0232.2-4420 & 42/35/55/58 & 1.78/7.08/5.01\\
RX J0528.9-3927 & 64/37/67/72 & 1.78/7.08/5.01\\
Zwicky 3146 & 111/78/128/140 & 1.12/4.47/3.16\\
RX J0043.4-2037 & 46/29/54/58 & 1.78/7.08/5.01\\
1E0657-56 & 267/196/309/340 & 0.50/2.24/1.41\\
Abell 2537$^{a}$ & 80/63/100/104 & 1.12/5.01/3.55\\
Abell 781 & 34/21/37/41 & 3.16/8.91/7.08\\
MACS J2245.0+2637 & 44/39/59/62 & 1.78/7.08/5.01\\
MACS J2311.5+0338 & 37/21/46/50 & 2.51/10.00/7.08\\
MACS J1131.8-1955 & 46/28/58/59 & 2.24/7.94/6.31\\
Abell 2744$^{a}$ & 120/98/150/159 & 1.26/5.01/3.55\\
MS2137.3-2353 & 54/44/66/68 & 0.56/2.82/1.78\\
MACS J0242.5-2132 & 28/20/36/37 & 2.51/8.91/6.31\\
Abell 1995 & 118/83/138/148 & 0.89/3.55/2.51\\
MACS J1427.6-2521 & 75/65/91/93 & 1.00/10.00/7.08\\
MACS J0547.0-3904 & 61/44/73/77 & 1.41/5.62/3.98\\
MACS J0257.6-2209 & 55/41/70/72 & 1.58/5.62/3.98\\

&& \\
\hline

\end{tabular*}
\end{table}

\addtocounter{table}{-1}

\begin{table}
\
\caption{\label{ChandraProperties}  Continued  }
\centering
\begin{tabular*}{0.9\columnwidth}{ c c c}\\
  \hline Cluster Name  & \mysub{N}{soft}/\mysub{N}{hard}/\mysub{N}{full}/\mysub{N}{any} &  Flux Limit  \\ 
\hline\hline

MACS J2049.9-3217 & 49/43/67/73 & 1.41/5.01/3.98\\
MACS J2229.7-2755 & 55/37/62/66 & 1.26/5.01/3.55\\
MACS J1319.9+7003 & 58/46/74/78 & 1.78/7.08/4.47\\
Zwicky J1358+6245 & 67/40/73/84 & 0.79/3.55/2.24\\
MACS J0520.7-1328 & 49/32/60/65 & 1.58/6.31/4.47\\
CL J0302-0423 & 28/25/38/42 & 2.51/11.22/7.08\\
MACS J1931.8-2634 & 104/92/133/137 & 0.63/2.51/1.78\\
MACS J0035.4-2015 & 46/40/59/61 & 1.58/5.62/3.98\\
CL J1212+2733 & 40/34/54/56 & 2.24/7.94/5.62\\
RBS 797 & 46/26/52/54 & 2.24/7.94/5.62\\
MACS J1115.8+0129 & 76/63/92/94 & 0.89/3.55/2.51\\
MACS J0308.9+2645 & 38/30/46/52 & 1.58/5.62/3.98\\
MACS J0404.6+1109 & 44/36/56/57 & 1.41/5.62/3.98\\
RX J0027.6+2616 & 23/19/32/36 & 2.51/10.00/7.08\\
RX J1532.9+3021$^{a}$ & 37/24/41/46 & 2.00/7.08/6.31\\
CL J0318-0302 & 43/31/51/54 & 2.00/7.94/5.62\\
Zwicky 1953 & 83/51/101/106 & 1.41/5.62/3.98\\
MACS J0011.7-1523 & 106/93/135/144 & 1.41/5.62/3.98\\
MACS J0949.8+1708 & 39/34/48/49 & 2.24/7.08/5.62\\
MACS J1720.2+3536 & 100/81/119/125 & 0.89/3.55/2.82\\
MACS J1731.6+2252 & 64/48/79/83 & 1.58/5.62/4.47\\
MACS J2211.7-0349 & 53/43/73/80 & 1.78/7.08/5.01\\
MACS J0429.6-0253 & 68/58/87/91 & 1.41/5.62/3.98\\
CL J0809+2811 & 36/26/47/52 & 1.78/7.08/5.01\\
V1416+4446 & 68/48/83/94 & 1.12/4.47/3.16\\
MACS J1006.9+3200 & 45/31/54/55 & 2.51/10.00/6.31\\
MACS J0159.8-0849 & 64/50/81/88 & 1.12/5.62/3.16\\
MACS J2228.5+2036 & 60/51/75/78 & 1.58/6.31/4.47\\
MACS J0152.5-2852 & 37/32/47/48 & 1.78/6.31/4.47\\
MACS J0159.0-3412 & 39/32/45/47 & 2.82/11.22/7.94\\
MACS J1105.7-1014 & 35/25/42/44 & 2.51/10.00/7.08\\
CL J1003+3253 & 56/39/66/69 & 1.78/6.31/4.47\\
MACS J2046.0-3430 & 72/64/89/91 & 0.89/3.55/2.51\\
MACS J0451.9+0006 & 39/31/50/52 & 2.82/10.00/7.08\\
MACS J0553.4-3342 & 23/18/35/37 & 2.82/10.00/7.08\\
MACS J0358.8-2955 & 24/23/36/37 & 3.16/11.22/7.94\\
MACS J1226.8+2153 & 119/111/154/76 & 1.78/6.31/4.47\\
MACS J1206.2-0847 & 59/53/84/86 & 1.41/5.01/3.98\\
CL J0141-3034 & 55/44/70/74 & 1.26/5.01/3.55\\
IRAS09104 & 22/18/29/32 & 2.51/11.22/7.08\\
MACS J0417.5-1154 & 100/87/125/131 & 1.00/4.47/4.47\\
MACS J2243.3-0935 & 48/38/58/62 & 1.78/5.62/4.47\\
MACS J0455.2+0657 & 41/23/48/49 & 2.51/10.00/7.08\\
MACS J1359.1-1929 & 34/23/43/44 & 2.51/10.00/7.08\\
MACS J0326.8-0043 & 25/15/34/36 & 2.51/10.00/7.08\\
MACS J0329.6-0211 & 74/62/94/102 & 0.89/3.55/2.51\\
RX J1347.5-1145 & 57/56/79/86 & 1.00/5.01/2.82\\
MACS J0140.0-0555 & 31/18/36/36 & 2.51/10.00/7.08\\
V1701+6414 & 63/51/80/84 & 1.00/3.55/2.51\\
3C295 & 122/101/145/153 & 0.56/2.51/1.58\\
MACS J1621.3+3810 & 134/112/154/159 & 0.71/2.82/2.00\\
CL J1641+4001 & 83/74/97/100 & 0.89/3.55/2.51\\
MACS J1115.2+5320 & 78/61/96/104 & 2.00/7.94/5.62\\
MACS J1108.8+0906 & 57/46/76/80 & 1.26/5.01/3.55\\
CL J0355-3741 & 42/34/50/56 & 1.41/5.62/3.55\\
CL J0333-2456 & 68/57/89/93 & 1.00/4.47/3.16\\
MACS J0111.5+0855 & 66/52/78/79 & 1.58/5.62/3.98\\

&& \\
\hline

\end{tabular*}
\end{table}

\addtocounter{table}{-1}

\begin{table}
\caption{\label{ChandraProperties}  Continued  }
\centering

\begin{tabular*}{0.9\columnwidth}{ c c c}\\
  \hline Cluster Name  & \mysub{N}{soft}/\mysub{N}{hard}/\mysub{N}{full}/\mysub{N}{any} &  Flux Limit  \\ 
\hline\hline

MACS J1427.2+4407 & 72/61/93/95 & 1.00/3.98/2.82\\
MACS J1311.0-0310 & 135/113/168/177 & 1.00/3.98/2.82\\
CL J1002+6858 & 41/39/57/62 & 1.58/6.31/4.47\\
RX J003033.2+261819 & 47/32/57/60 & 1.78/7.08/5.01\\
MACS J2214.9-1359 & 79/58/99/104 & 1.12/4.47/3.16\\
MACS J0911.2+1746 & 71/49/81/88 & 1.00/3.98/2.82\\
MACS J0257.1-2325 & 78/56/95/106 & 1.26/5.01/3.55\\
V1525+0958 & 71/52/87/94 & 1.00/3.16/2.51\\
CL J1357+6232 & 70/61/93/97 & 0.89/3.55/2.51\\
MACS J0454.1-0300$^{a}$ & 66/45/75/85 & 2.24/7.94/6.31\\
MACS J1423.8+2404$^{a}$ & 103/83/121/127 & 1.41/7.08/4.47\\
MACS J1149.5+2223 & 80/65/99/107 & 2.24/7.94/6.31\\
MACS J0717.5+3745 & 127/92/141/156 & 1.00/3.98/2.82\\
MS0015.9+1609 & 90/78/109/116 & 0.63/2.51/1.78\\
V1121+2327 & 108/90/131/135 & 0.63/2.51/1.78\\
CL J0216-1747 & 98/74/114/123 & 0.89/5.01/3.55\\
MACS J0025.4-1222 & 142/111/168/182 & 0.89/4.47/2.82\\
CL J0956+4107 & 90/65/112/118 & 1.00/3.98/2.82\\
MACS J2129.4-0741 & 85/67/101/106 & 1.12/4.47/3.16\\
CL0328-2140 & 71/66/95/100 & 0.89/3.55/2.51\\
MACS J0647.7+7015 & 68/50/89/91 & 1.12/4.47/3.16\\
CL J1120+4318 & 56/38/70/74 & 1.58/6.31/4.47\\
CL J1334+5031 & 49/45/70/70 & 2.00/7.08/5.01\\
CL J0542.8-4100 & 113/72/127/136 & 0.89/3.16/2.24\\
CL J1202+5751 & 77/72/102/108 & 0.79/3.16/2.24\\
CL J0405-4100 & 82/68/101/112 & 1.26/5.62/3.98\\
MACS J0744.8+3927 & 104/95/130/139 & 0.71/3.16/2.24\\
V1221+4918 & 103/95/139/149 & 0.63/2.51/1.78\\
CL J0152.7-1357 & 95/65/116/121 & 1.12/3.98/2.82\\
CL J1226.9+3332$^{a}$ & 109/84/128/131 & 1.00/4.47/3.55\\

&& \\
\hline

\end{tabular*}
\end{table}

\section{Results on Cluster AGN Counts and Spatial Distribution}
Our final point source catalogs and sensitivity maps determine the distribution of X-ray bright AGN across the cluster fields. In all cases, the dominant uncertainty in our measurements is the Poisson uncertainty in the measured number of sources. The expected Poisson fluctuations for samples of size $n$ are estimated using the 1-$\sigma$ asymmetric confidence limits of \citep{Gehrels1986}.  The 1-$\sigma$ upper confidence limit $\lambda_{U}$ and 1-$\sigma$ lower confidence limit $\lambda_{L}$ for a sample of $n$ sources are approximated as  

\begin{eqnarray*}
\lambda_{U}(n)=n + 1 + \sqrt{0.75 +n} \\ 
\lambda_{L}(n)=n\left(1-\frac{1}{9n}-\frac{1}{3\sqrt{n}}\right)^{3}
\end{eqnarray*}

\noindent These estimates are accurate to within a few percent for all values of $n$. Monte Carlo simulations estimating the impact of uncertainties in the source flux measurements and systematic uncertainties in the sensitivity maps show that these uncertainties are negligible compared to the Poisson uncertainties on the source counts. 

We have compared our results on the X-ray point source population in the cluster fields to both the \cdfs \  and the \cha \ {\it COSMOS} deep field surveys \citep{Elvis2009,Puccetti2009}. In order to minimize the effects of differences in the analysis pipelines and calibration products used, the {\it COSMOS} results presented here are the result of a re-analysis of those data using our pipeline and with measurements made in the same energy bands. The analysis of the \cdfs \ fields used a pipeline similar to ours, and no re-analysis was required to enable a direct comparison. Unless otherwise noted, all numbers and fluxes listed correspond to the full band catalog, and have been corrected for Galactic absorption.

\subsection{Cumulative Number Counts}

\begin{figure*}
\centering
\subfigure[]{
\includegraphics[width=0.49\textwidth, angle=270]{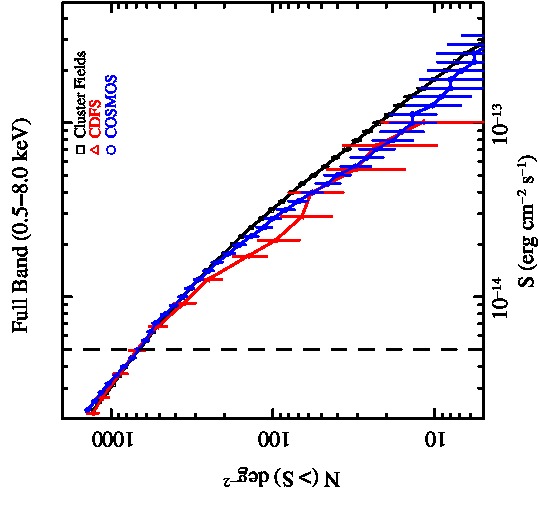}
\label{SoftlogNlogS}
}
\subfigure[]{
\includegraphics[width=0.47\textwidth, angle=270]{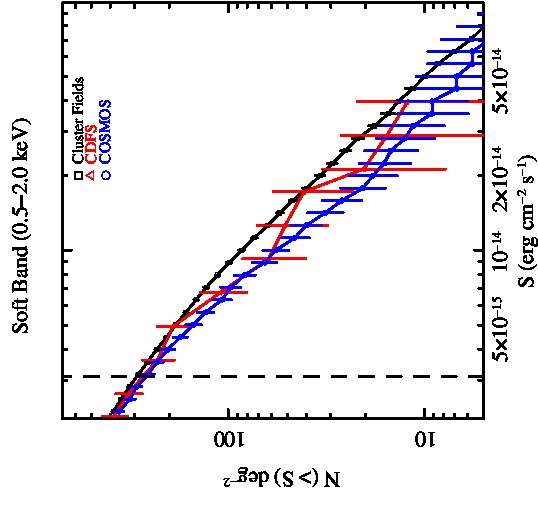}
\label{HardlogNlogS}
}
\subfigure[]{
\includegraphics[width=0.47\textwidth, angle=270]{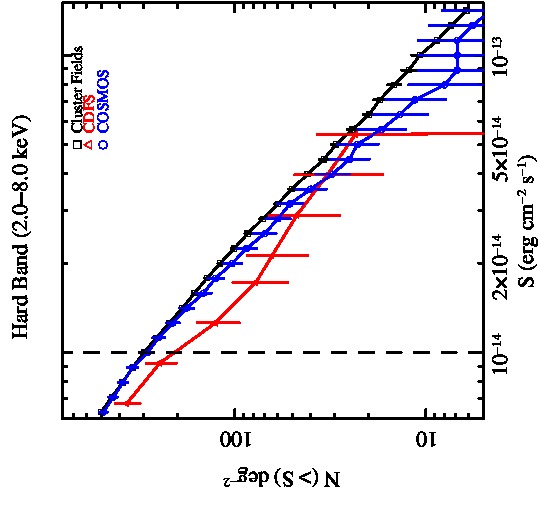}
\label{FulllogNlogS}
}

\caption{\label{NumberCounts}  Cumulative number counts ($\log{N}-\log{S}$) in the full ($0.5-8.0 \keV$, a), soft ($0.5-2.0 \keV$, b), and hard ($2.0-8.0 \keV$, c) energy bands for the cluster fields (black). The red curves show the cumulative number counts in the same energy bands for the \cdfs.  The blue curves are the results from the {\it COSMOS} survey. The band-specific flux limits used to determine the radial distribution of X-ray point sources are denoted by the vertical dashed line in each figure. In all three bands an excess of sources at fluxes $\gtrsim 10^{-14} \ergpcmsqps$ with respect to the control fields is observed.  }

\end{figure*}

The cumulative number density of sources above a given flux ($S$) is calculated as 

\begin{equation}
N (> S) = \sum_{S_{i} > S} \frac{1}{\Omega_{i}}
\end{equation}

\noindent where $\Omega_{i}$ is the total survey area sensitive to the $i^{th}$ source flux $S_{i}$. The $\log{N}-\log{S}$ cumulative number counts for sources in the full, soft, and hard energy bands are shown in Figure \ref{NumberCounts}, together with a comparison to the \cdfs \ and {\it COSMOS} results in the same bands \citep{Lehmer2012}. The cumulative number counts for the cluster and field sources show the commonly observed broken power-law shape \citep{Cowie2002,Moretti2003,Bauer2004,Lehmer2012}. Above fluxes of $\sim 10^{-14} \ergpcmsqps$, the cluster fields exhibit a slight excess in source density compared to field surveys. These results are consistent with and build on those discussed in Paper I, and demonstrate the robustness of this analysis procedure.

\subsection{The Radial Distribution of X-ray Sources}\label{RadProfSec}

The spatial distribution of point sources about the cluster centers has been calculated for all point sources with full-band fluxes above $1 \times 10^{-14} \ergpcmsqps$. Similar analyses were performed in the soft band and hard bands, with flux limits of $3 \times 10^{-15} \ergpcmsqps$ and $10^{-14} \ergpcmsqps$, respectively. The full band flux limit corresponds to a luminosity of $\sim 10^{42} \ergps$ for the lowest redshift cluster in this sample (Abell 2163) and $\sim 10^{43} \ergps$ for the highest redshift cluster (CL J1226.9+3332). 

\begin{figure*}
\centering
\subfigure[]{
\includegraphics[width=0.49\textwidth, angle=270]{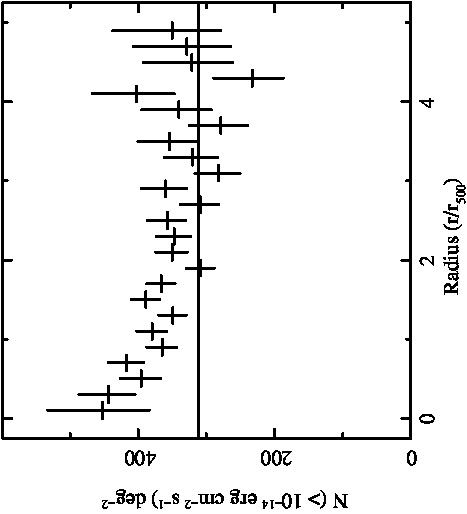}
\label{RadProfSoft}
}
\subfigure[]{
\includegraphics[width=0.47\textwidth, angle=270]{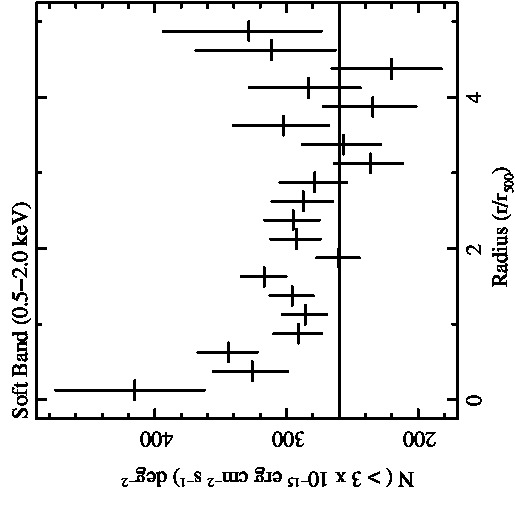}
\label{RadProfHard}
}
\subfigure[]{
\includegraphics[width=0.47\textwidth, angle=270]{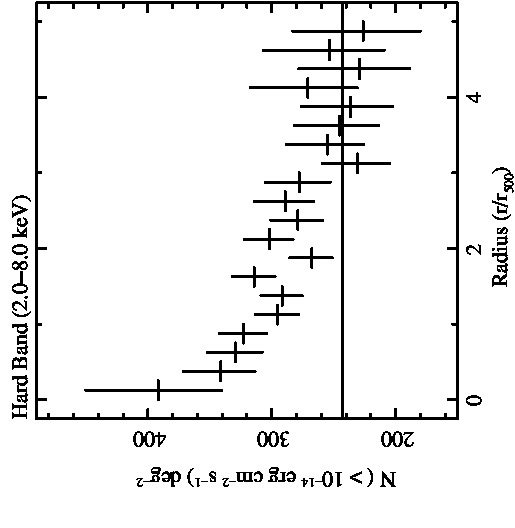}
\label{RadProfFull}
}

\caption{\label{RadProf}  The projected density of X-ray bright point sources in all three bands, in units of $\rm{deg}^{-2}$. In all three lines, the solid black line corresponds to the best-fit constant background density in the range 3-5 \rfive, and in all three cases this background density is consistent with the expected field source density derived from \cdfs \ and {\it COSMOS}. In all three energy bands, this constant background field density is consistent with the expected field density determined from the \cdfs \ and {\it COSMOS} data. (a): The surface density of X-ray bright full band sources ($\mysub{F}{X}(0.5-8.0 \keV) > 1 \times 10^{-14} \ergpcmsqps$) as a function of radius, in units of \rfive.  A total of 2675 sources were included in the calculation of this profile. (b): The surface density of X-ray bright soft band ($\mysub{F}{X}(0.5-2.0 \keV) > 3 \times 10^{-15} \ergpcmsqps$) sources as a function of radius, in units of \rfive. A total of 3055 sources were included in the calculation of this profile.  (c): The surface density of X-ray bright hard band sources ($\mysub{F}{X}(2.0-8.0 \keV) > 10^{-14} \ergpcmsqps$) as a function of radius, in units of \rfive. A total of 2933 sources were included in the calculation of this profile.}

\end{figure*}

The adoption of these flux limits minimizes complications due to residual incompleteness and systematic uncertainties in the sensitivity maps, while maintaining a strong statistical signal. A total of 6443, 3055, and 2933 sources satisfy these criteria in the full, soft, and hard bands, respectively. The projected radial distributions are plotted in Figure \ref{RadProf} as a function of radius in units of \rfive. The projected radii of sources in each cluster field were calculated assuming that they lie at the cluster redshift. The projected source density profile and its statistical uncertainties in each radial bin are calculated in an identical manner to that used to calculate the cumulative number counts.

In this representation, we find clear evidence for an excess of point sources in the central regions of the clusters. At large radii, the measured source number densities converge to an approximately constant source density. Fitting the number density of full ($0.5-8.0 \keV$) band sources between 3-5 \rfive \ with a constant model provides an estimated background number density of $311 \pm 16 \ \rm{deg}^{-2}$.\footnote{The constant model provides a statistically acceptable fit to the data ($\chi^{2}=4.7$ for $\nu=7$ degrees of freedom).} The measured value is also in agreement with the expected background source density from the \cdfs \ and {\it COSMOS} studies within statistical uncertainties. Within the projected central virialized cluster region ($\sim 2 $\rfive), the constant background density model provides a poor fit to the point source density, and can be rejected at $ > 99.9\%$ confidence. The results of the background fits in all three bands are shown in Table \ref{RadProfResults}. The high statistical precision of our data enable us to measure an excess of approximately 3 sources per cluster field within 2 \rfive \ in each energy band.  We do not expect any significant contribution to this signal from gravitational lensing given the results of \citep{Refregier1997} and \cite{Gilmour2009}. In fact, given the shape of the cumulative number counts ($\log{N}-\log{S}$), gravitational lensing is expected to suppress the detection of sources near the centers of clusters \citep{Refregier1997,Gilmour2009}.

\begin{table*}
\caption{\label{RadProfResults} Flux limits for the radial profiles and the expectations for the X-ray point source density from \cdfs \ and {\it COSMOS} deep fields in all three energy bands. The columns list: (1) the energy band; (2) the flux limit used in constructing the radial profile fits, in units of \ergpcmsqps; (3) the measured background density between 3 and 5 \rfive \ from the radial profile;  (4) the number of sources detected within 2\rfive \ at that flux limit, across all clusters; (5) the survey area within 2\rfive \ at that flux limit, in units of $\deg^{-2}$; (6) the average excess number of sources per cluster above that flux limit within 2\rfive, determined by extrapolating measurements of the field density from the best-fit constant model between 3 and 5 \rfive;  (7) the best-fit power-law index for the projected source density of cluster member AGN; (8) the density of field sources from the \cdfs \ at that flux limit;  and (9) the density of field sources from {\it COSMOS} at that flux limit.   }
\centering

\begin{tabular}{ c c c c c c c c c }\\
  \hline 

(1) & (2) & (3) & (4) & (5) & (6) & (7) & (8) & (9)  \\
Band & Flux Limit  & Cluster Fields ($\deg^{-2}$) & $\mysub{n}{2}$ & $\mysub{\Omega}{2}$ ($\deg^{2}$)& Excess  &  $\beta$ & \cdfs ($\deg^{-2}$) & {\it COSMOS} ($\deg^{-2}$)  \\ 
\hline\hline
Full & $1 \times 10^{-14}$   & $311 \pm 16$  & 2474  & 6.6 & $3.1 \pm 0.5$ & $-0.42 \pm 0.12$ & $330 \pm 48$ & $356 \pm 21$  \\
Soft & $3 \times 10^{-15}$   & $263 \pm 14$  & 2683 & 8.8 & $2.8 \pm 0.4$ & $-0.53 \pm 0.22$  & $250 \pm 45$ & $255 \pm 18$  \\
Hard & $1 \times 10^{-14}$   & $244 \pm 14$  & 2595 & 8.5 & $3.1 \pm 0.4$ & $-0.48 \pm 0.14$  & $220 \pm 45$ & $287 \pm 19$  \\
&&&&&&&\\
\hline\hline

\end{tabular}
\end{table*}

We have fitted the observed X-ray point source density profiles in all three bands with a King-law+Constant model:  

\begin{equation}\label{Powerlaw}
\mysub{N}{X}(r)=\frac{\mysub{N}{0}}{1+\left(\frac{r}{\mysub{r}{c}}\right)^{2}} + \mysub{C}{X}
\end{equation}

\noindent where $\mysub{r}{c}$ is the core radius of the fit. The resulting posterior distributions for the fits in each energy band are nearly identical to one another. In each case, we measure a median core radius of $\mysub{r}{c}=1.2$ \rfive, with a $68\%$ confidence interval spanning the range of $\mysub{r}{c} \in [0.7,2.1]$ \rfive. Most published studies of the optical galaxy population in clusters measure the projected galaxy density profile to follow a King Model or NFW model with a scale radius of $\sim 0.2-0.5$ \rfive \citep{Popesso2007,Budzynski2012}.  King models with core radii $\mysub{r}{c} < 0.5$ \rfive \ can be rejected at $\gtrsim 99\%$ confidence. This indicates that the fraction of cluster member galaxies hosting X-ray AGN rises with radius (see also Paper II). Fitting the observed X-ray point source density profile to a power-law model ($\mysub{N}{X}(r) \sim r^{\beta}$) gives similar results as in Paper I: we measure a median power-law index of $\beta =-0.5 \pm 0.1$ consistently across all three energy bands. 

\subsubsection{The Distribution of Luminous Cluster Member AGN}
We have also determined the radial distribution of X-ray point sources above the field using a full band luminosity limit of $L \geq 3 \times 10^{43} \ergps$ after a statistical subtraction of the field population. For each cluster we determined the flux limit corresponding to $L = 3 \times 10^{43} \ergps$ at the cluster redshift, and then calculated for each radial bin the number of sources detected and number of expected field sources\footnote{We use our determinations of the {\it COSMOS} $\log{N}-\log{S}$ to determine the number of sources expected from the field in each radial bin.} brighter than that flux limit. The projected number density of excess sources above this luminosity limit is given by the difference of these two values in each radial bin, divided by the total survey area. We use Monte Carlo simulations to determine the error bars on each of these measurements. 

Our calculations show that these luminous AGN are distributed out to distances of $\sim 2.5 \rfive$, beyond which the excess number density is consistent with zero. Fitting this profile to a power-law model provides a best-fit logarithmic slope of $-0.5 < \beta < -0.6$, which is consistent with the power-law slope measured for the flux limited sample without statistical field subtraction. The measured excess corresponds to a total of $\sim 1 $ excess sources with $\mysub{L}{X} \geq 3 \times 10^{43} \ergps$ per cluster.

 \begin{figure}
\centering
\includegraphics[width=0.92\columnwidth, angle=270]{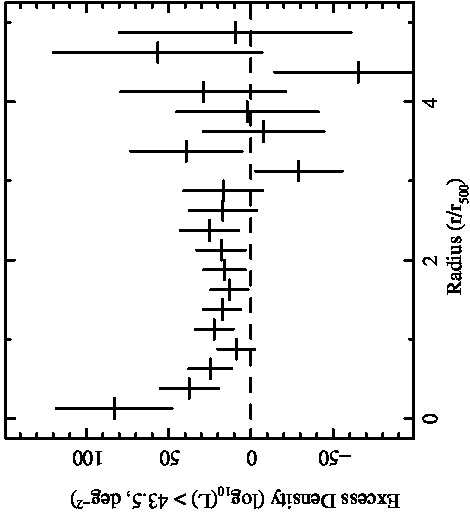}
\label{ExcessLumDens}
\caption{\label{ExcessL435} The projected density of X-ray point sources detected above a full band luminosity limit of $L \geq 3 \times 10^{43} \ergps$, in units of $\rm{deg}^{-2}$. This projected source density follows the same power-law model as that observed for the flux-limited sample.   }
\end{figure}

\begin{figure}
\centering
\includegraphics[width=0.92\columnwidth]{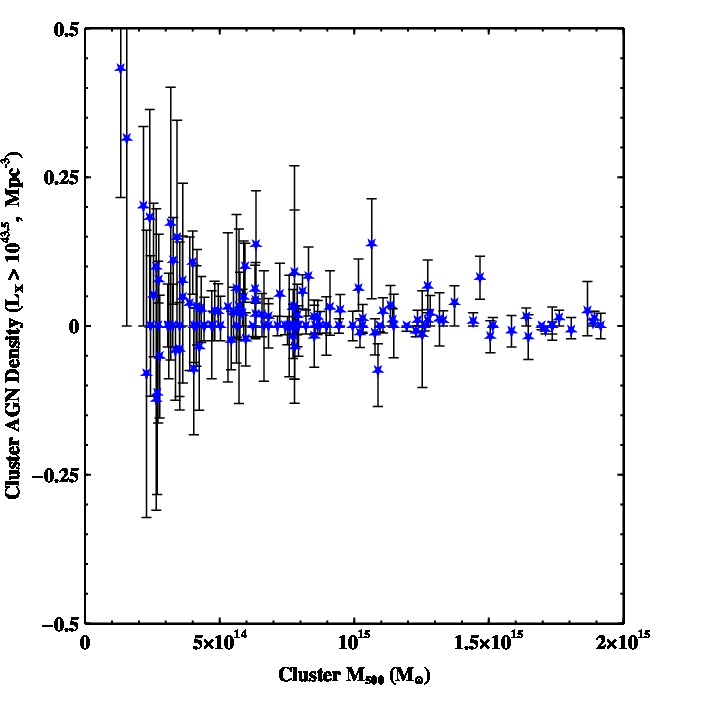}
\caption{ \label{ExcessbyMass}The comoving number density of X-ray AGN more luminous than $3 \times 10^{43} \ergps$ within $2\rfive$ for our cluster sample, as a function of cluster mass $\mysub{M}{500}$. These number densities were determined by statistically subtracting the expected number of field sources in each cluster aperture using the field AGN density as determined by {\it COSMOS} from the number of sources we detect; hence negative AGN densities are possible. While only a few of the clusters have excesses that are individually larger than zero with high statistical significance, there is nevertheless evidence that lower mass clusters host larger AGN densities within $2\rfive$ as compared to more massive clusters. }
\end{figure}

Using this same luminosity limit, we also determined the comoving number density of cluster member X-ray AGN within $2\rfive$ in each of our 135 galaxy clusters in Figure \ref{ExcessbyMass} after statistical field subtraction. Althouhg the statistical significance of any individual cluster's excess AGN density is small, there is nevertheless evidence for lower overall number densities of AGN in more massive clusters. 

\section{Testing Mass and Redshift Dependent Models}

Such a large and well-characterized sample allows us to measure the specific evolution of cluster AGN versus that in the field. To this end, we utilize a Markov Chain Monte Carlo (MCMC) analysis procedure to determine posterior probability distributions for parameters in a redshift, luminosity, and cluster mass dependent model for the projected point source density profile.  We confront our model with the data from each of the 135 galaxy clusters, marginalizing over uncertainties in the expected evolution of X-ray AGN in the field and the density of background sources expected in our survey.\footnote{By background sources, we mean X-ray point sources coincident with the cluster along the line of sight that are not at the cluster redshift. These sources have been shown to have a roughly constant density across survey areas as large as $\sim 1 \deg^{2}$ \citep{Xue2011,Elvis2009}. } More specifically, our model assumes that the projected number density of cluster sources (in units of $\deg^{-2}$) above a given flux limit $f$, at a redshift $z$ and projected distance $r$ from the center of a cluster of mass $\mysub{M}{500}$, is proportional to the co-moving number density of X-ray AGN in the field at the cluster redshift (known as the X-ray Luminosity Function or XLF) with a power-law spatial dependence:
\begin{equation}
\mysub{N}{obs}(>f, r, z)=  \mysub{N}{} \times  \mysub{D}{A}(z)^{2} \times \rfive \times \Phi(>\mysub{L}{cut}, z) \times  \left(\frac{r}{\rfive}\right)^{\beta} + C
\end{equation}
\noindent where $\Phi(>\mysub{L}{cut}, z)$ is the expected co-moving number density (in units of $\Mpc^{-3}$) of X-ray AGN at that redshift in the luminosity range of $\mysub{L}{cut} < L < 10^{46} \ergps$ as determined by the XLF model of \cite{Ueda2014}. The lower limit of the luminosity function \mysub{L}{cut} is the intrinsic luminosity of an AGN at the cluster redshift corresponding to the survey flux cut-off $f$ in the survey of $10^{-14} \ergpcmsqps$. This flux cut-off corresponds to a luminosity range of $\sim 10^{42-43} \ergps$ for the cluster member AGN.  We assume that the cluster AGN contribution arises within a cylinder, centered on the cluster, whose line-of-sight depth scales with \rfive. $\mysub{D}{A}(z)$ is the angular diameter distance specific to each cluster. The parameter $\mysub{N}{}$ includes the necessary unit conversions and describes the factor by which the number density of AGN in clusters exceeds the field value specified by the XLF (hereafter the scaling factor). $C$ is the (constant) density of field AGN at our flux limit $f$.  We allow the scaling factor to vary as a power law in mass and redshift 
\begin{equation}
\mysub{N}{} \rightarrow \mysub{N}{0} (1+z)^{\eta} \left(\frac{\mysub{M}{500}}{10^{15} \msolar}\right)^{\zeta}
\end{equation}

 \noindent and also allow the radial distribution to depend linearly on the cluster mass and redshift as 
\begin{equation}
 \beta \rightarrow \mysub{\beta}{0} + \mysub{\beta}{z} (1+z) + \mysub{\beta}{m} \left(\frac{\mysub{M}{500}}{10^{15} \msolar }\right)
\end{equation}
\noindent Our null hypothesis is that the AGN population in clusters scales with the expected field behavior (i.e. the cluster AGN population evolves in a scaled manner with respect to the field AGN population across all redshifts and for clusters of all masses), which in terms of our model means that all mass and redshift dependent terms (i.e. $\zeta, \eta, \mysub{\beta}{z} \ \& \ \mysub{\beta}{m}$) should be statistically consistent with $0$. 

Our MCMC analysis provides several key advantages over a more traditional statistical analysis, in particular: 1) It uses the full information of cluster redshifts and masses without the need to resort to binning; 2) we are able to interpret the results within the context of our complex selection function, which varies the overall luminosity limit from cluster to cluster as well the area in each radial bin sensitive to sources of a given flux; and 3) we are able to determine robustly the covariances between the different model parameters, which are difficult to anticipate {\it a priori}.

\subsubsection{The XLF Model}
Before presenting the results from our MCMC runs, it is important to discuss the choice of XLF for this study in more detail. 
For this study, we assume the Luminosity-Dependent Density Evolution (LDDE) XLF model of \cite{Ueda2014}. The XLF of \cite{Ueda2014} was determined in the rest frame $2-10 \keV$ band, while we are using the $0.5 -8.0 \keV$ band in order to maximize the statistics of our measurement. In order to account for this energy band conversion, we convert the relevant parameters of the \cite{Ueda2014} model ($\mysub{L}{\star}, \mysub{L}{a_{1}} \& \ \mysub{L}{a_{2}}$) to the full band assuming a power-law photon index of $\Gamma=1.4$. Additionally, we allow our priors to have statistical uncertainties a factor of 2 larger than the error bars published in \cite{Ueda2014}, in order to account for the fact that the XLF may take on slightly different shapes in these two energy bands. However, the majority of the parameters for this model of the XLF are consistent with those measured in softer energy bands \citep{Hasinger2005}, suggesting that this procedure should not introduce any significant systematic error in our analysis. We also emphasize that this model is almost identical to the model of \cite{Ueda2003} at the redshifts of the clusters. 

This luminosity function takes a double power-law form, parameterized as:

\begin{equation}
\frac{\rm{d} \Phi(\mysub{L}{X},0)}{\mathrm{d} \log{\mysub{L}{X}}}= \frac{\mysub{A}{0}}{\left[\mysub{L}{X}/L^{*}\right]^{\mysub{\gamma}{1}}+\left[\mysub{L}{X}/L^{*}\right]^{\mysub{\gamma}{2}}}
\end{equation}

\noindent where $\frac{\rm{d} \Phi(\mysub{L}{X},0)}{\mathrm{d} \log{\mysub{L}{X}}}$ is the differential number density of X-ray AGN (in units of $\Mpc^{-3}$ per logarithmic unit of luminosity) at redshift $z=0$. For the LDDE model we have chosen, redshift evolution in the luminosity function is parameterized as 
\begin{equation}
\frac{\rm{d} \Phi(\mysub{L}{X},z)}{\mathrm{d} \log{\mysub{L}{X}}}=\frac{\rm{d} \Phi(\mysub{L}{X},0)}{\mathrm{d} \log{\mysub{L}{X}}} e(z)
\end{equation}
\noindent with a redshift correction factor $e(z)$ that takes the form of

 \begin{displaymath}
   e(z) = \left\{
     \begin{array}{lr}
       (1+z)^{\mysub{p}{1}} & : z \leq \mysub{z}{c_{1}}\\
       \left(1+\mysub{z}{c_{1}}\right)^{\mysub{p}{1}} \left(\frac{1+z}{1+\mysub{z}{c_{1}}}\right)^{\mysub{p}{2}} & : \mysub{z}{c_{1}} < z \leq \mysub{z}{c_{2}}\\
       \left(1+\mysub{z}{c_{1}}\right)^{\mysub{p}{1}} \left(\frac{1+z}{1+\mysub{z}{c_{1}}}\right)^{\mysub{p}{2}} \left(\frac{1+z}{1+\mysub{z}{c_{2}}}\right)^{\mysub{p}{3}} & : z > \mysub{z}{c_{2}}
     \end{array}
   \right.
\end{displaymath} 
\noindent where \mysub{z}{c_{1}} \& \mysub{z}{c_{2}} are the two transition redshifts between the different evolution indexes ($\mysub{p}{1}, \mysub{p}{2}, \& \  \mysub{p}{3}$). In the LDDE model, the transition redshifts also depend on luminosity as  
\begin{displaymath}
   \mysub{z}{c_{1}}(\mysub{L}{x}) = \left\{
     \begin{array}{lr}
       \mysub{z}{c_{1}}^{*} & : \mysub{L}{X} > \mysub{L}{a_{1}}\\
       \mysub{z}{c_{1}}^{*} \left(\frac{\mysub{L}{X}}{\mysub{L}{a_{1}}}\right)^{\mysub{\alpha}{1}} & : \mysub{L}{X} < \mysub{L}{a_{1}}
     \end{array}
   \right.
\end{displaymath} 

\noindent and similarly for \mysub{z}{c_{2}} 
\begin{displaymath}
   \mysub{z}{c_{2}}(\mysub{L}{x}) = \left\{
     \begin{array}{lr}
       \mysub{z}{c_{2}}^{*} & : \mysub{L}{X} > \mysub{L}{a_{2}}\\
       \mysub{z}{c_{2}}^{*} \left(\frac{\mysub{L}{X}}{\mysub{L}{a_{2}}}\right)^{\mysub{\alpha}{2}} & : \mysub{L}{X} < \mysub{L}{a_{2}}
     \end{array}
   \right.
\end{displaymath}

\noindent Finally, the first evolution index $\mysub{p}{1}$ scales with luminosity as 

\begin{equation}
\mysub{p}{1}(\mysub{L}{X})=\mysub{p}{1}^{*} + \mysub{\beta}{1} \times (\log{\mysub{L}{X}}-44)
\end{equation}

The full set of parameter values utilized for our study including their priors are found in Table \ref{Priors}.

\begin{table*}
\caption{\label{Priors} Input priors on the MCMC runs of our four models. Nearly all of these priors are determined by the measurements of the XLF  after converting published results to our energy band. All of the priors with error bars shown are assumed to be normally distributed, while those without error bars are fixed. Our priors have error bars a factor of $2$ larger than the published values in order to account for any potential systematics that may arise in the energy band conversion. The only additional prior included in our analysis is for $C$, the projected density of X-ray AGN in the field as determined by {\it COSMOS}, assumed to be normally distributed with a variance of $10\%$, which is sufficiently large to account for both the statistical fluctuations and cosmic variance in this measurement.     }
\centering

\begin{tabular}{ c c c }\\
  \hline 
{\bf XLF Priors} \\
Parameter & Prior  \\ 
\hline\hline
$\mysub{A}{0} \ (\Mpc^{-3} \ \rm{dex}^{-1}) $ & $(2.91 \pm 0.14) \times 10^{-6} $ \\     
$\mysub{\gamma}{1}$ & $0.96 \pm 0.08$ \\
$\mysub{\gamma}{2}$ & $2.71 \pm 0.18$ \\
$\log{\mysub{L}{\star}}$ & $43.97 \pm 0.12$ \\
$\mysub{p}{1}^{*}$ & $4.78 \pm 0.16$ \\
$\mysub{p}{2} $ & $-1.5$ \\
$\mysub{p}{3} $ & $-6.2$ \\
$\mysub{z}{c_{1}}^{*}$ & $1.86 \pm 0.14$ \\
$\mysub{z}{c_{2}}^{*}$ & $3.0$ \\
 $\mysub{\beta}{1} $ & $0.84 \pm 0.36$ \\
$\log{\mysub{L}{a_{1}}} $ & $44.61 \pm 0.14$ \\
$\log{\mysub{L}{a_{2}}} $ & $44.00$ \\  
$\mysub{\alpha}{1}$ & $0.29 \pm 0.04$\\ 
$\mysub{\alpha}{2}$ & $-0.1$\\
$C \ (\deg^{-2}$) & $330 \pm 33$ \\ 
\hline\hline

\end{tabular}
\end{table*}

\section{Results on Cluster Mass and Redshift Dependent Parameters }

\begin{figure*}
\centering
\subfigure[]{
\includegraphics[width=0.55\textwidth]{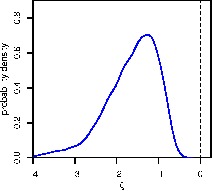}
\label{Model1}
}
\subfigure[]{
\includegraphics[width=0.47\textwidth]{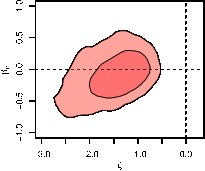}
\label{Model2}
}
\subfigure[]{
\includegraphics[width=0.47\textwidth]{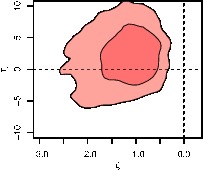}
\label{Model3}
}

\caption{\label{ModelConfs} Posterior confidence intervals for model parameters. {\it Top:} The 1-dimensional posterior probability distribution for $\zeta$ in Model 2, where $\zeta$ is the only model parameter that is not fixed to its null value of $0$. The null hypothesis of $\zeta=0$ (denoted by the dashed vertical line) can be rejected at $> 99.9\%$ confidence. {\it Bottom Left:} The two-dimensional confidence contours ($68.3\% \ \& \ 95.4\%$) for Model 2, where $\zeta$ and $\mysub{\beta}{m}$ are both free parameters. The null hypotheses of $\mysub{\beta}{m}, \zeta=0$ are denoted by the dashed lines. This model provides a consistent value for $\zeta$ as Model 2 and demonstrates that the mass dependence of Model 2 is inconsistent with arising from a mass-dependence in the spatial distribution of the cluster AGN. {\it Bottom Right:} The two-dimensional confidence contours ($68.3\% \ \& \ 95.4\%$) for Model 3, where $\zeta$ and $\eta$ are free parameters. The null hypotheses of $\eta, \zeta=0$ are denoted by the dashed lines. This model provides a consistent value for $\zeta$ as Models 1 and 2 and demonstrates that the mass dependent scaling factor we observe is inconsistent with a model with a redshift dependence beyond the expected field evolution.}
\end{figure*}

We find most parameters in our model to be consistent with their respective null hypotheses. For example, there is no significant evidence in our data for the spatial distribution of AGN in galaxy clusters to vary in a statistically significant manner with either the cluster redshift or mass (i.e. $\mysub{\beta}{z}$ \& $\mysub{\beta}{m}$ are consistent with $0$). The scaling factor does not have any significant redshift dependence (i.e. $\eta \sim 0$). However, the scaling factor does appear to have a strong mass dependence (i.e. $N \sim M^{\zeta}$): the null hypothesis of $\zeta=0$ can be rejected at high ($\gtrsim 99.9\%$) confidence. With all four parameters of interest (i.e. $\zeta, \eta, \mysub{\beta}{z}, \& \mysub{\beta}{m}$) free (hereafter Model 1), the mode (i.e. the peak of the posterior distribution) is located at approximately $\zeta \sim -1.6$, with a $68.3\%$ confidence interval of $\sim [-2.1,-0.9]$. We have further confirmed the robustness of this result by comparing the posterior distributions of two independent sub-samples of clusters, each of which gives consistent measurements for the posterior distributions of all four parameters. This result provides clear evidence for peculiar evolution of X-ray AGN in galaxy clusters beyond the expectations from the field. To explore this result further, we present the results of posterior probability distributions from three additional models, the results of which are shown in Figure \ref{ModelConfs}: Model 2 fixes all of the cluster mass and redshift dependent terms to $0$ with the exception of $\zeta$; Model 3 includes both $\zeta$ and $\mysub{\beta}{m}$ as free parameters; while Model 4 allows $\zeta$ and $\eta$ to be free parameters. Model 2 provides the most precise constraints on the mass dependent evolution of the scaling factor. Model 3 demonstrates that the mass-dependent scaling factor we measure is not degenerate with a dependence in the spatial distribution of the cluster AGN. Model 4 shows that our sample of cluster AGN evolves with redshift in a manner consistent with expectations from the field, and also that we have the statistical power to distinguish between mass-dependent and redshift-dependent models. Other models that freeze and thaw different combinations of these four parameters have also been examined, and the results of those models are consistent with the four models presented in this text.

The 1-dimensional posterior probability distributions along with the priors are summarized in Table \ref{ModelPars}. In Table \ref{ModelPars}, we list the mode and the $68\%$ confidence interval about that mode for each free parameter. The posterior probability distributions are typically non-Gaussian in shape, and often have long asymmetric tails extending beyond their modes. Our sample provides little to no constraint regarding the redshift dependence of the scaling factor ($\eta$). The data are also consistent with a redshift and mass-independent radial profile for the cluster X-ray AGN. Our most constraining model for a mass dependent scaling factor constrains the value of that power-law slope to $\zeta \in [-3.71,-0.60]$ for its 99\% confidence interval.

\begin{table*}
\caption{\label{ModelPars} The resulting parameter values from our posterior probability distributions for all four models. For each parameter, we show the mode and the $68\%$ confidence interval about that value, as determined by the 1-dimensional posterior probability distributions.  We find that only one parameter shows statistically significant deviations from our field prediction: $\zeta$, the power-law dependence of the scaling factor with mass. }

\centering

\begin{tabular}{ c c c c c}\\
{\bf Posteriors} \\
Parameter & Model 1 & Model 2 & Model 3  & Model 4\\ 
\hline\hline
\smallskip
\smallskip
\smallskip
$\eta$ &  $1.90^{+2.2}_{-1.8}$ &$0 $ &  $0$ & $0.97^{+3.03}_{-2.89}$ \\
\smallskip
\smallskip
$\zeta$ & $-1.63^{+0.50}_{-0.75}$ & $-1.18^{+0.30}_{-0.80}$  & $-1.28^{+0.35}_{-0.45}$ & $-1.32^{+0.34}_{-0.51}$   \\
\smallskip
\smallskip
$\mysub{\beta}{0}$ & $-1.55^{+1.30}_{-2.00}$ & $-0.63^{+0.18}_{-0.14}$ & $-0.67^{+0.21}_{-0.16}$ & $-0.67 \pm 0.14$   \\
\smallskip
\smallskip
$\mysub{\beta}{z}$ & $0.78^{+1.40}_{-1.00}$ & $0$ & $0$ & $0$   \\
\smallskip
\smallskip
$\mysub{\beta}{m}$ & $-0.18^{+0.35}_{-0.20}$& $0$ & $-0.03^{+0.16}_{-0.28}$ & $0$    \\
\smallskip
\smallskip
$C \ (\deg^{-2}$) & $350^{+12}_{-12}$ & $345^{+14}_{-20}$ & $342^{+11}_{-17}$ & $335^{+14}_{-18}$   \\
\hline\hline

\end{tabular}
\end{table*}

\subsection{Physical Interpretation}

The most straightforward interpretation of the $M^{-1.2}$ scaling relation that we observe is that it is driven by galaxy mergers within the cluster. Using virial arguments, we expect that the galaxy velocity dispersion, $\sigma$, in clusters will scale with cluster mass as $\sim M^{1/3}$. Additionally, theoretical calculations suggest that the rate of mergers between cluster galaxies should scale as $\sim \sigma^{-3}$ \citep{Mamon1992}, or equivalently $\sim M^{-1}$, consistent with the $ M^{\sim -1.2}$ scaling observed in these data (Model 2).

Other physical explanations beyond a merger-driven scenario may possibly result in the observed number density profiles for these data. Any alternative model, however, would have to provide consistent results for all four mass and redshift dependent parameters. The absence of any dependence in the spatial distribution of these AGN with cluster mass and redshift suggests that galaxy-ICM interactions such as ram pressure stripping are not responsible for driving this scaling relation: it is unlikely that these process would operate on the same length scales (in units of \rfive) irrespective of mass yet lower the overall scaling factor so noticeably, especially since the physical length of \rfive in each cluster scales with mass. Further simulation work of cluster galaxies falling through realistic ICM environments from large distances will be necessary to further investigate these possibilities, however.

\section{Preliminary Spectral Identification and Visual Classification of Cluster Member AGN}

We have carried out a preliminary attempt to confirm cluster member AGN spectroscopically by searching the NASA/IPAC Extragalactic Database (NED) for optical spectroscopic counterparts for our X-ray point source positions. The search circle around each X-ray source is $2 \arcsec$, sufficiently large to account for the expected positional uncertainties on our X-ray sources. Control tests that added random offsets to the X-ray source positions suggest that our expected number of ``false positives'' (i.e. finding a spectroscopic counterpart at the cluster redshift by chance coincidence) is negligible. Where we find a spectroscopic counterpart to the X-ray source with a redshift $\mysub{z}{cp}$ satisfying $c\vert \mysub{z}{cluster}-\mysub{z}{cp} \vert < 5000 \km \s^{-1}$, we identify that X-ray source to be a spectroscopically confirmed cluster member. In total, we find that 88 of our X-ray AGN have spectroscopic counterparts within $2\arcsec$ of the source position.

We then searched the {\it Hubble} archive for images at each of these source positions made with either the ACS or WFC3 cameras. The {\it Hubble} images were registered to the \cha \ images and cleaned of cosmic rays using the Laplacian edge detection algorithm of \cite{VanDokkum2001}. After these steps, 23 of the X-ray AGN had {\it Hubble} images deemed suitable for a preliminary visual classification of their morphologies. 

Source catalogs for each {\it Hubble} field were produced using \sext \ \citep{Bertin1996} in a single filter. For each X-ray AGN we selected three control galaxies with similar optical magnitudes and clustercentric distances to the X-ray AGN. We then produced postage stamp images of the 5$\arcsec$ radius surrounding each galaxy in both the AGN and control sample, utilizing up to three filters of imaging data for each galaxy when available. The postage stamp {\it Hubble} images for all 23 X-ray AGN can be found in Figure \ref{XRaySourceImages}. Information about the filters and source positions are given in Table \ref{XRaySourceTab}.

\begin{table*}
\caption{\label{XRaySourceTab}  Information about the {\it Hubble} images for all 23 spectroscopically confirmed X-ray AGN. For each source position, we denote the filters used for the images presented in Figure \ref{XRaySourceImages}. }
\centering

\begin{tabular}{ c c c c c  }\\
  \hline 
RA & DEC & Filter 1 & Filter 2  & Filter 3   \\ 
\hline\hline

3.53939 & -30.41137 &  F606W &  &  \\ 
3.55994 & -30.37781 &  F606W &  &  \\ 
3.61061 & -30.39563 &  F606W &  &  \\ 
4.63350 & 16.49064 &  F606W &  &  \\ 
4.65420 & 16.46027 &  F775W & F606W & F555W \\ 
6.38361 & -12.38467 &  F555W &  &  \\ 
28.16550 & -13.92369 &  F775W & F625W &  \\ 
28.28761 & -13.96692 &  CLEAR1L &  &  \\ 
28.30403 & -13.89717 &  F606W &  &  \\ 
29.99430 & -8.82704 &  F606W &  &  \\ 
73.50511 & 2.96277 &  F606W &  &  \\ 
73.55489 & 2.95945 &  F606W &  &  \\ 
139.43829 & 51.71885 &  CLEAR1L &  &  \\ 
146.80429 & 76.38735 &  F125W & F110W & F606W \\ 
151.72157 & 32.00314 &  F606W &  &  \\ 
177.39105 & 22.37405 &  F775W & F606W & F475W \\ 
181.55492 & -8.79565 &  F775W & F606W & F475W \\ 
186.71535 & 21.87390 &  F606W &  &  \\ 
186.75880 & 33.56825 &  F775W & F625W &  \\ 
212.63080 & 52.25929 &  F775W &  &  \\ 
224.31136 & 22.32598 &  F850LP & F775W &  \\ 
243.90677 & -6.18664 &  F606W &  &  \\ 
328.38098 & 17.69271 &  F125W & F850LP &  \\ 
\hline\hline

\end{tabular}
\end{table*}

Galaxy morphologies were determined visually to fall into one of the following classes: 1) {\bf Disturbed} galaxies which have clear signatures of disruptions from mergers such as tidal tails; 2) {\bf Undisturbed}  galaxies with no apparent disruptions; 3) Nearby {\bf Neighbor} galaxies which, while not having evidence for major disruptions, are sufficiently near to other galaxies to suggest an imminent merger; 4) {\bf Stellar} galaxies whose morphologies could not be distinguished from a point source; or 5) {\bf Empty} images where the host of the X-ray point source could not be determined. All of the co-authors except authors SE, RC, and AvdL did the morphology classification on all 92 galaxies. None of the participating co-authors knew which of the galaxies were the hosts of the X-ray AGN and which were control galaxies in advance. We then determined the fraction of galaxies within each of these morphological classes for both the normal galaxies and X-ray AGN.

Our main finding is that galaxies hosting X-ray AGN were classified as {\bf Disturbed} at higher rates than the control sample. The {\bf Disturbed} vote fractions for the X-ray AGN and control sample are $22\% $ (30/138) and $10\% $ (43/414), respectively. We utilize a two-sided Student's t-test to determine the probability of these two measurements arising from the same underlying population. Assuming a null hypothesis where the X-ray AGN and control samples have the same average fractions of disturbed galaxies, then we have a probability of $p=0.043$ of measuring a larger absolute difference in the disturbed fractions of the two samples.\footnote{This particular $p$ value assumes equal variance for the X-ray AGN and control samples. If we don't assume equal variance, then the probability of measuring a larger absolute difference in the means increases to $p=0.097$.} Galaxies hosting X-ray AGN were equally likely to be classified as {\bf Neighbor, Empty, Stellar}, and {\bf Undisturbed} as their control galaxies. Applying a simple weighting scheme that weights more strongly sources where all voters agree on a particular classification leads to similar results. In conclusion, while there is some initial evidence that X-ray AGN in clusters may be preferentially hosted in galaxies with disturbed morphologies (consistent with a merger driven scenario), larger data samples and more robust (preferably automated) classification schemes will be required to further investigate the extent to which mergers may be responsible for the triggering of X-ray AGN in galaxy clusters.  

One particular galaxy of note identified in this study is the X-ray AGN at $\rm{RA (J2000)}= 04^{h} 54^{m} 13.17^{s}$, $ \rm{DEC (J2000)}=+02^{\circ} 57^{m} 34.0^{s}$ in the galaxy cluster Abell 520, a cluster galaxy which hosts a clear partial Einstein ring that has not yet been published in the literature.

\clearpage

 \begin{figure*}
 \centering

\includegraphics[width=0.37\textwidth,natwidth=541,natheight=579]{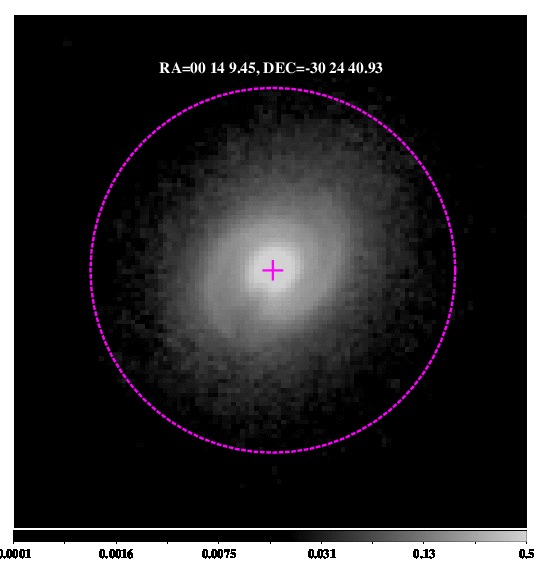}
\includegraphics[width=0.37\textwidth,natwidth=541,natheight=579]{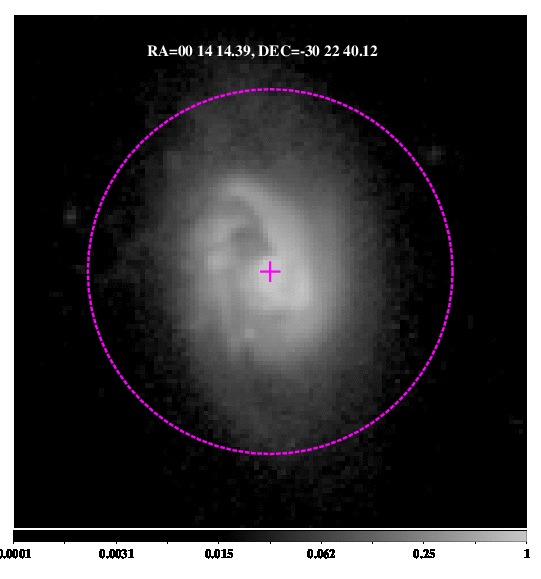}
\includegraphics[width=0.37\textwidth,natwidth=541,natheight=579]{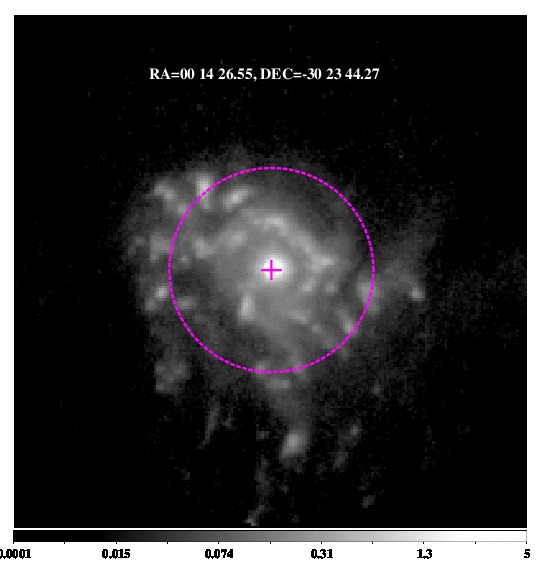}
\includegraphics[width=0.37\textwidth,natwidth=541,natheight=579]{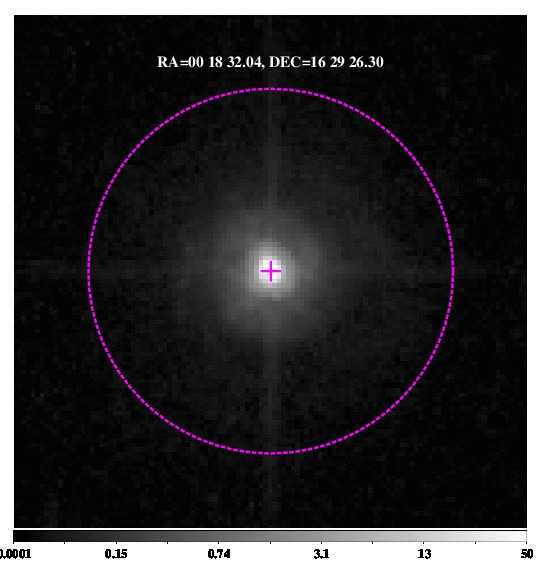}
\includegraphics[width=0.37\textwidth,natwidth=541,natheight=579]{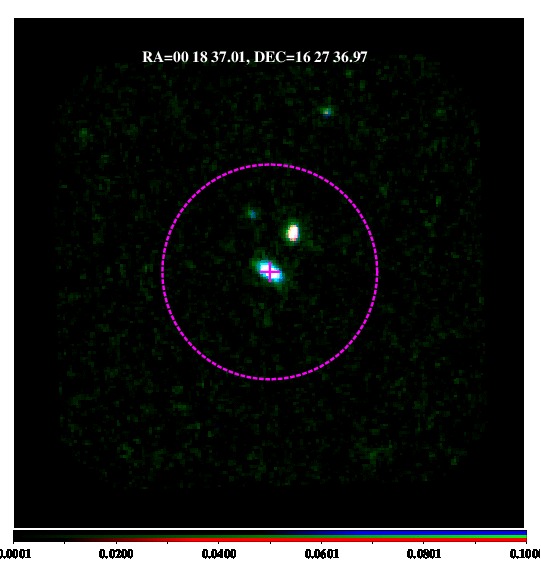}
\includegraphics[width=0.37\textwidth,natwidth=541,natheight=579]{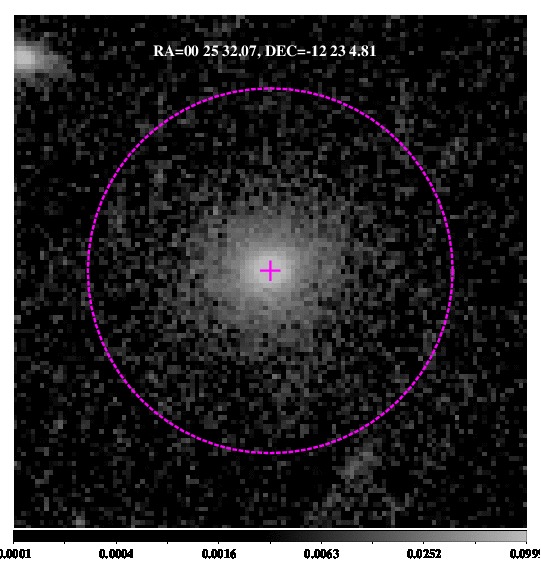}

\caption{\label{XRaySourceImages}  The 23 spectroscopically confirmed X-ray images in our sample with follow-up {\it Hubble} imaging. The cross in each image denotes the position of the X-ray source, and the dashed circle corresponds to a radius of 2\arcsec \ surrounding the AGN, which is roughly a factor of two larger than the positional uncertainty for all of our X-ray sources.        }
\centering

 \end{figure*}

\addtocounter{figure}{-1}

 \begin{figure*}
 \centering

\includegraphics[width=0.37\textwidth,natwidth=541,natheight=579]{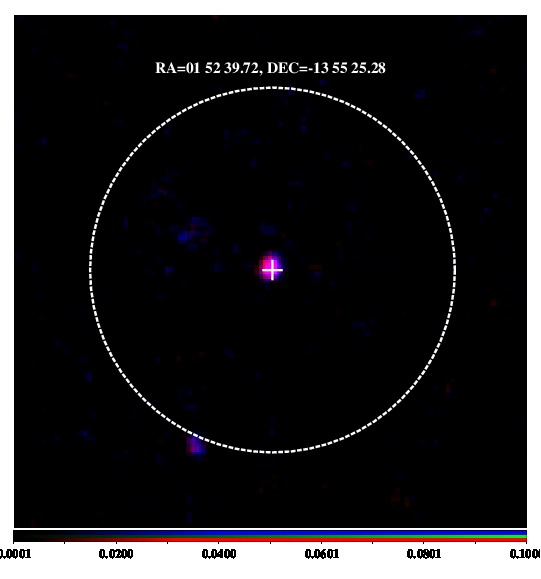}
\includegraphics[width=0.37\textwidth,natwidth=541,natheight=579]{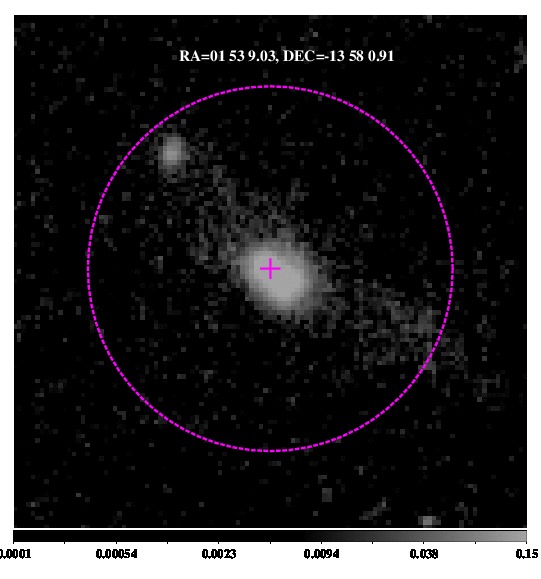}
\includegraphics[width=0.37\textwidth,natwidth=541,natheight=579]{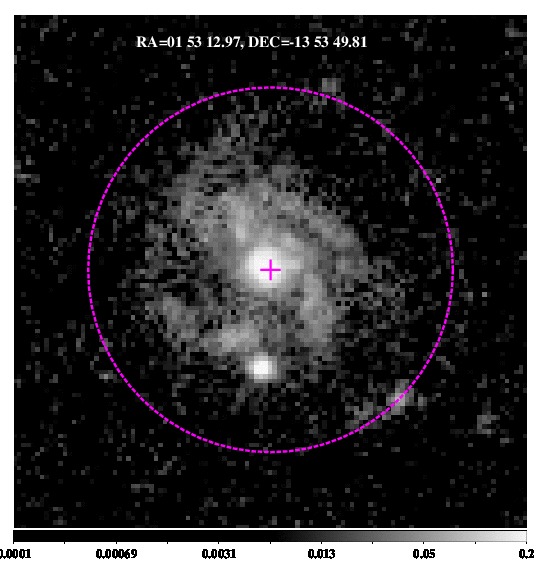}
\includegraphics[width=0.37\textwidth,natwidth=541,natheight=579]{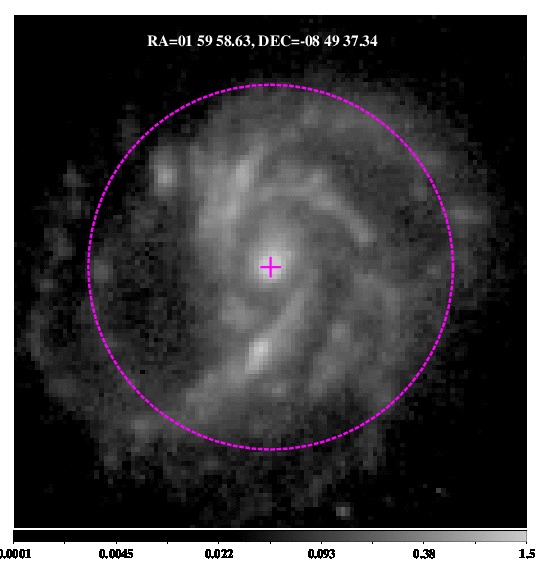}
\includegraphics[width=0.37\textwidth,natwidth=541,natheight=579]{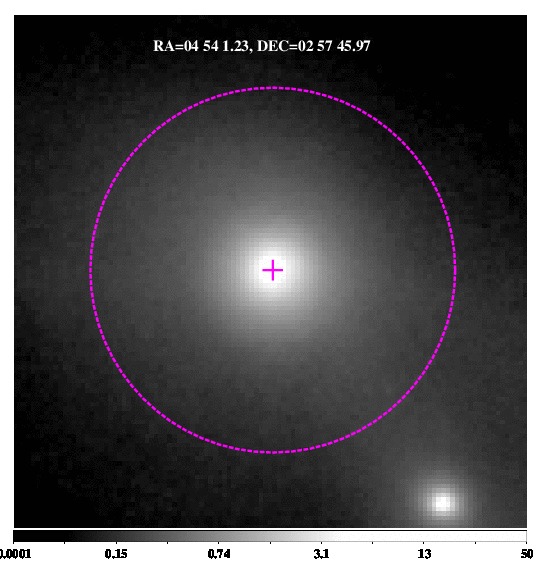}
\includegraphics[width=0.37\textwidth,natwidth=541,natheight=579]{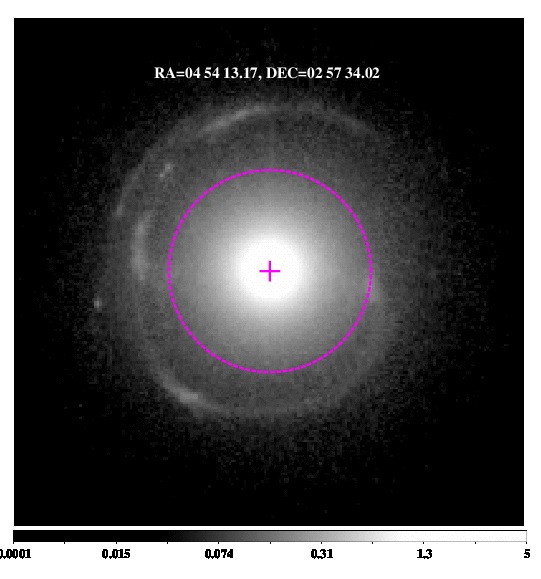}

\caption{\label{XRaySourceImages}  Continued. The bottom right image in this panel is a galaxy with a previously unpublished partial Einstein ring.           }

 \end{figure*}

\addtocounter{figure}{-1}

 \begin{figure*}
 \centering

\includegraphics[width=0.37\textwidth,natwidth=541,natheight=579]{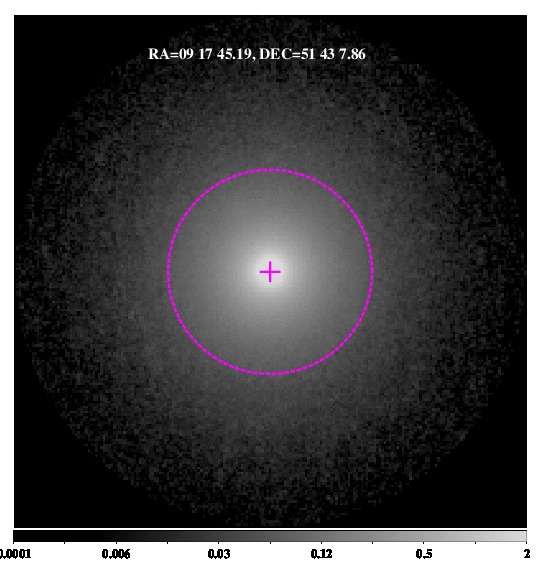}
\includegraphics[width=0.37\textwidth,natwidth=541,natheight=579]{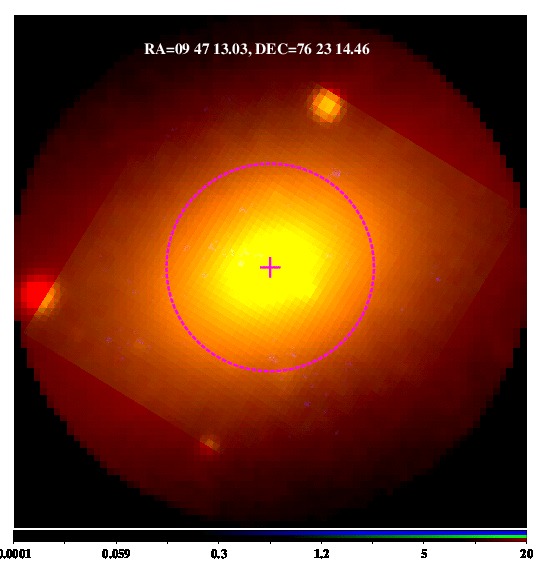}
\includegraphics[width=0.37\textwidth,natwidth=541,natheight=579]{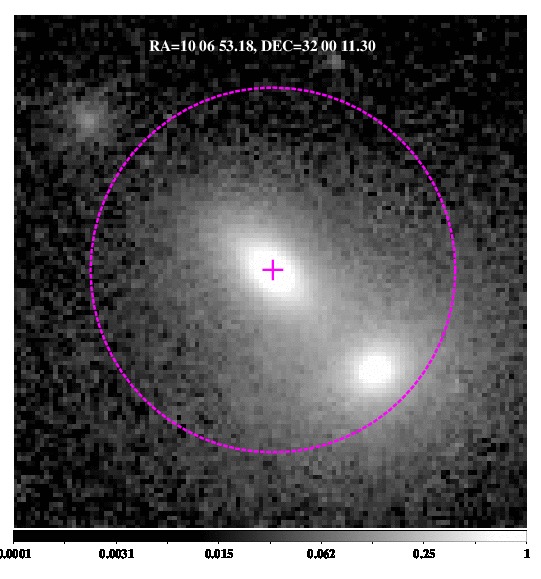}
\includegraphics[width=0.37\textwidth,natwidth=541,natheight=579]{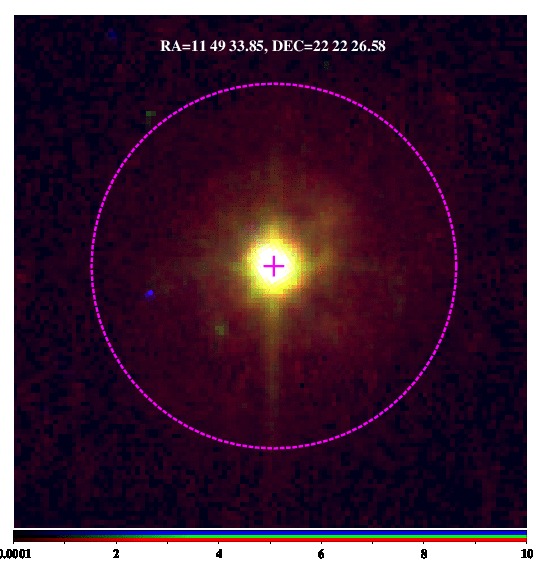}
\includegraphics[width=0.37\textwidth,natwidth=541,natheight=579]{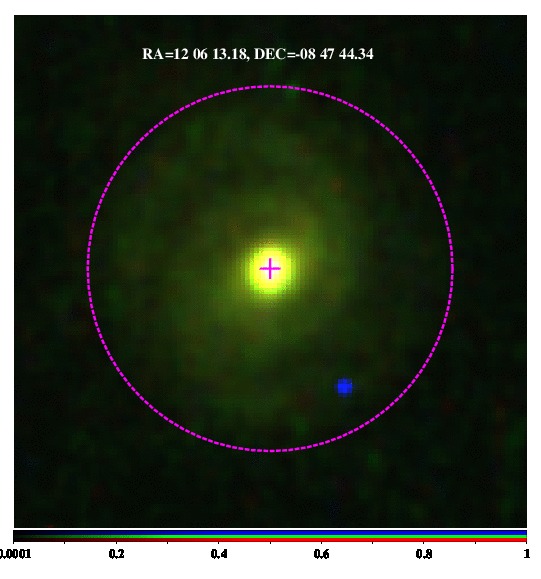}
\includegraphics[width=0.37\textwidth,natwidth=541,natheight=579]{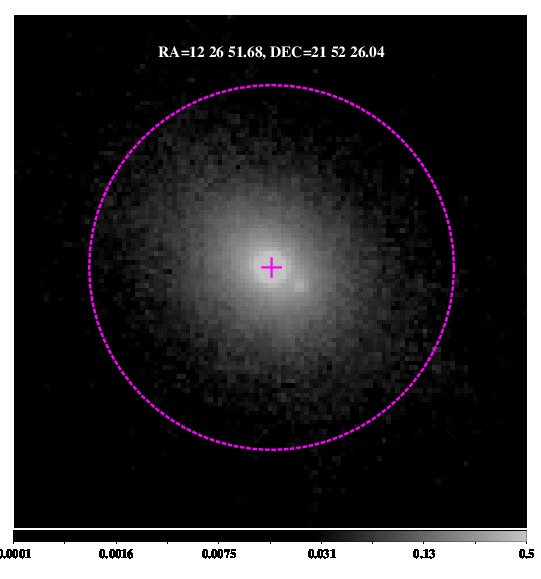}

\caption{\label{XRaySourceImages}  Continued        }

 \end{figure*}

\addtocounter{figure}{-1}

 \begin{figure*}
 \centering

\includegraphics[width=0.37\textwidth,natwidth=541,natheight=579]{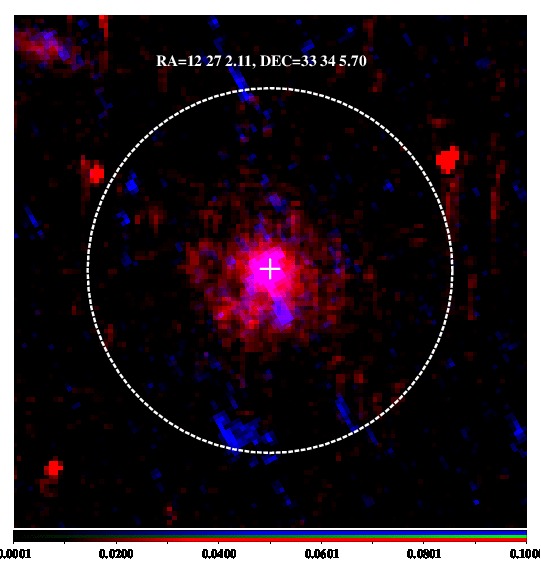}
\includegraphics[width=0.37\textwidth,natwidth=541,natheight=579]{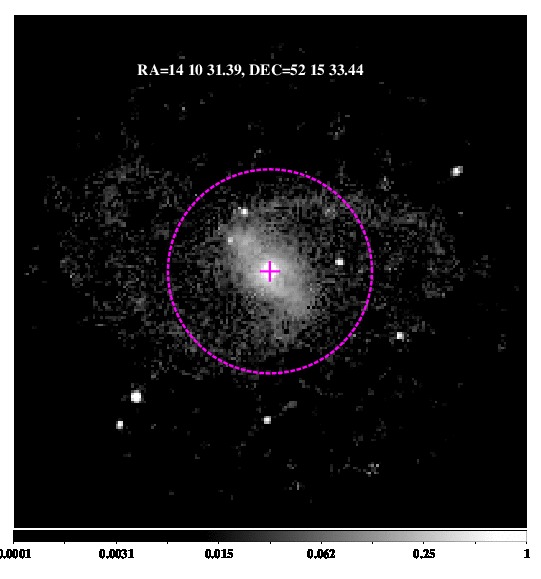}
\includegraphics[width=0.37\textwidth,natwidth=541,natheight=579]{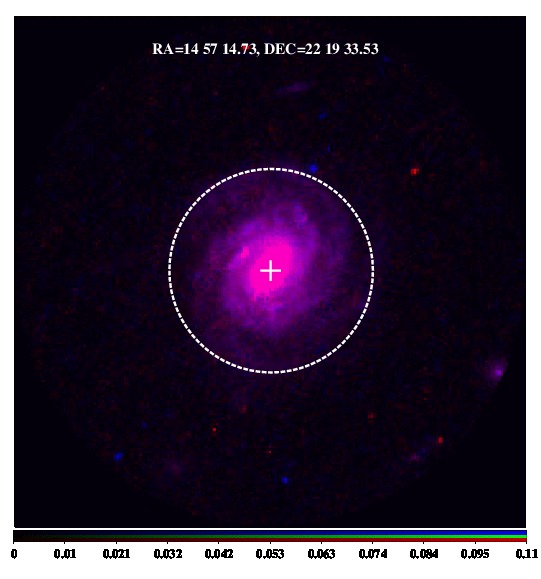}
\includegraphics[width=0.37\textwidth,natwidth=541,natheight=579]{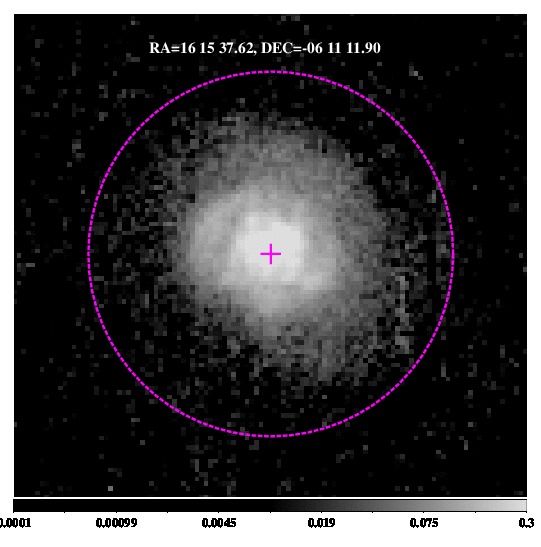}
\includegraphics[width=0.37\textwidth,natwidth=541,natheight=579]{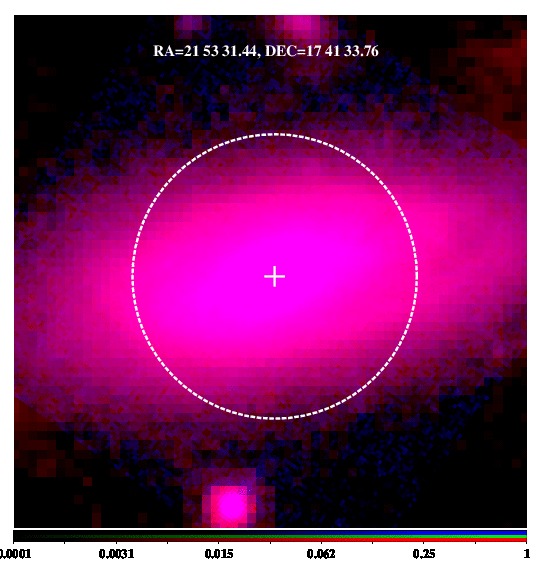}

\caption{\label{XRaySourceImages}  Continued        }

 \end{figure*}

\section{Discussion}

A merger-driven scenario for the triggering of X-ray AGN in clusters would be interesting in the context of recent literature results. Investigations into the morphologies of X-ray AGN host galaxies in the field at similar X-ray luminosities, stellar masses, and redshifts to the sample presented here have found no connection between galaxy morphologies and the presence of an X-ray AGN \citep[e.g.][]{Gabor2009,Cisternas2011,Schawinski2011,Kocevski2012,Schawinski2012,Simmons2012,Fan2014}. Our results therefore suggest that AGN in galaxy clusters may be triggered by distinct physical processes to those of field AGN. Further studies with larger samples of both cluster and field AGN will be necessary to understand the origins of this apparent dichotomy.

The $\sim M^{-1}$ scaling relation that we find may also play an important role in understanding the evolution of the AGN fraction in galaxy clusters over cosmic time. Previous work has demonstrated that the X-ray AGN fraction in this galaxy cluster sample is $\sim 3$ times lower than in the field (Paper II), while studies of lower mass (mean masses of $\mysub{M}{500} \lesssim 10^{14}$), higher redshift clusters and groups ($z \sim 1$) appear to indicate higher fractions of AGN than in the field \citep{Lehmer2013, Martini2013}. The origin of the turnaround between these high redshift, low mass systems and lower redshift, more massive clusters may be driven by the fact that galaxy clusters grow continuously over cosmic time (and subsequently acquire higher velocity dispersions leading to lower merger rates).

Larger samples of galaxy clusters observed with \cha \ and {\it Hubble} will be required to further investigate the origin of this signal, as all of the measurements here are limited by the sample statistics and not any systematic uncertainties in the XLF. Larger samples will be essential to place stronger constraints on the redshift dependent terms ($\eta, \mysub{\beta}{z}$), as these terms in particular are not well constrained by these data. Recent Sunyaev-Zeldovich surveys of galaxy clusters such as those from the South Pole Telescope or Atacama Cosmology Telescope will be especially useful for similar studies as they offer more leverage for high redshift clusters than the studies utilized here. The morphological comparisons between cluster member X-ray AGN and control galaxies is also limited by the number statistics of spectroscopically confirmed X-ray AGN with both \cha \ and {\it Hubble} imaging data.

From the theoretical side, more accurate calculations of the merger rate of galaxies in clusters will be required to further examine the scenario we propose. Our hypothesis that cluster AGN are driven by galaxy mergers hinges on the analytic prediction of the cluster galaxy merger rate in \cite{Mamon1992}, and we have not found literature results that discuss the accuracy of this prediction using N-body or hydrodynamic simulations of realistic galaxy clusters. Indeed, it remains uncertain whether the cluster galaxy merger rate scales with the cluster properties (such as redshift) or galaxy-specific properties (such as stellar mass or clustercentric distance) in ways unaccounted for in this analytic prediction. Confirming or refuting this prediction for the galaxy merger rate in clusters will provide critical information as to the interpretation of these results, and may also be able to provide other possible origins for the $\sim M^{-1.2}$ scaling relation we measure.

\section*{Acknowledgments}

We thank Patrick Broos and Leisa Townsley for technical support, insight, and advice as to how to utilize best the \ae \ software package for this study. We also thank Bret Lehmer for providing the cumulative number counts data for the \cdfs \ and Nina Bonaventura at the \cha \ X-ray Center's Helpdesk for technical support in the catalog production. Support for this work was provided by the Department of Energy Grant Number DE-AC02-76SF00515 and \cha \ X-ray Center grant GO0-11149X. Support for SE  was provided by the Smithsonian Astrophysical Observatory (SAO) subcontract SV2-82023 under NASA contract NAS8-03060. 
We also acknowledge support from NASA ADP grant NNX10AC99G, \cha \ X-ray Center grants SP1-12007A, and National Science Foundation grants AST-0838187 \& AST-1140019. We also acknowledge support from the Thousand Young Talents (QingNianQianRen) program (KJ2030220004), the USTC startup funding (ZC9850290195), and the National Natural Science Foundation of China through NSFC-11243008. Additional funding as provided by the Strategic Priority Research Program ``The Emergence of Cosmological Structures'' of the Chinese Academy of Sciences (XDB09000000).  This research has made use of the NASA/IPAC Extragalactic Database (NED) which is operated by the Jet Propulsion Laboratory, California Institute of Technology, under contract with the National Aeronautics and Space Administration. Based on observations made with the NASA/ESA Hubble Space Telescope, and obtained from the Hubble Legacy Archive, which is a collaboration between the Space Telescope Science Institute (STScI/NASA), the Space Telescope European Coordinating Facility (ST-ECF/ESA) and the Canadian Astronomy Data Centre (CADC/NRC/CSA).

\bibliographystyle{mnras}
\def \aap {A\&A} 
\def \statisci {Statis. Sci.}
\def \physrep {Phys. Rep.}
\def \pre {Phys.\ Rev.\ E}
\def \sjos {Scand. J. Statis.} 
\def \jrssb {J. Roy. Statist. Soc. B} 


%

\def \araa {ARA\&A}
\def \aj {AJ}
 \def \aas {A\&AS}
\def \apj {ApJ}
\def \apjl {ApJL}
\def \apjs {ApJS}
 \def \aaps {Ap\&SS}
\def \mnras {MNRAS}
\def \nat {Nat}
 \def \pasp {PASP}
\def \gca {Geochim.\ Cosmochim.\ Acta}
\def \prd {Phys.\ Rev.\ D}
\def \prl {Phys.\ Rev.\ Lett.}

\bibliography{AllRefs1}

\begin{thebibliography}{}

\bibitem[\protect\citeauthoryear{{Allen}, {Evrard} \& {Mantz}}{{Allen}
  et~al.}{2011}]{Allen2011}
{Allen} S.~W.,  {Evrard} A.~E.,    {Mantz} A.~B.,  2011, \araa, 49, 409

\bibitem[\protect\citeauthoryear{{Applegate}, {von der Linden}, {Kelly},
  {Allen}, {Allen}, {Burchat}, {Burke}, {Ebeling}, {Mantz} \&
  {Morris}}{{Applegate} et~al.}{2014}]{Applegate2012}
{Applegate} D.~E.,  {von der Linden} A.,  {Kelly} P.~L.,  {Allen} M.~T.,
  {Allen} S.~W.,  {Burchat} P.~R.,  {Burke} D.~L.,  {Ebeling} H.,  {Mantz} A.,
    {Morris} R.~G.,  2014, \mnras, 439, 48

\bibitem[\protect\citeauthoryear{{Bauer}, {Alexander}, {Brandt}, {Schneider},
  {Treister}, {Hornschemeier} \& {Garmire}}{{Bauer} et~al.}{2004}]{Bauer2004}
{Bauer} F.~E.,  {Alexander} D.~M.,  {Brandt} W.~N.,  {Schneider} D.~P.,
  {Treister} E.,  {Hornschemeier} A.~E.,    {Garmire} G.~P.,  2004, \aj, 128,
  2048

\bibitem[\protect\citeauthoryear{{Bertin} \& {Arnouts}}{{Bertin} \&
  {Arnouts}}{1996}]{Bertin1996}
{Bertin} E.,  {Arnouts} S.,  1996, \aaps, 117, 393

\bibitem[\protect\citeauthoryear{{B{\"o}hringer}, {Schuecker}, {Guzzo},
  {Collins}, {Voges}, {Cruddace}, {Ortiz-Gil}, {Chincarini}, {De Grandi},
  {Edge}, {MacGillivray}, {Neumann}, {Schindler} \& {Shaver}}{{B{\"o}hringer}
  et~al.}{2004}]{Bohringer2004}
{B{\"o}hringer} H.,  {Schuecker} P.,  {Guzzo} L.,  {Collins} C.~A.,  {Voges}
  W.,  {Cruddace} R.~G.,  {Ortiz-Gil} A.,  {Chincarini} G.,  {De Grandi} S.,
  {Edge} A.~C.,  {MacGillivray} H.~T.,  {Neumann} D.~M.,  {Schindler} S.,
  {Shaver} P.,  2004, \aap, 425, 367

\bibitem[\protect\citeauthoryear{{Brandt} \& {Hasinger}}{{Brandt} \&
  {Hasinger}}{2005}]{Brandt2005}
{Brandt} W.~N.,  {Hasinger} G.,  2005, \araa, 43, 827

\bibitem[\protect\citeauthoryear{{Broos}, {Townsley}, {Feigelson}, {Getman},
  {Bauer} \& {Garmire}}{{Broos} et~al.}{2010}]{Broos2010}
{Broos} P.~S.,  {Townsley} L.~K.,  {Feigelson} E.~D.,  {Getman} K.~V.,  {Bauer}
  F.~E.,    {Garmire} G.~P.,  2010, \apj, 714, 1582

\bibitem[\protect\citeauthoryear{{Budzynski}, {Koposov}, {McCarthy}, {McGee} \&
  {Belokurov}}{{Budzynski} et~al.}{2012}]{Budzynski2012}
{Budzynski} J.~M.,  {Koposov} S.,  {McCarthy} I.~G.,  {McGee} S.~L.,
  {Belokurov} V.,  2012, ArXiv e-prints

\bibitem[\protect\citeauthoryear{{Burenin}, {Vikhlinin}, {Hornstrup},
  {Ebeling}, {Quintana} \& {Mescheryakov}}{{Burenin}
  et~al.}{2007}]{Burenin2007}
{Burenin} R.~A.,  {Vikhlinin} A.,  {Hornstrup} A.,  {Ebeling} H.,  {Quintana}
  H.,    {Mescheryakov} A.,  2007, \apjs, 172, 561

\bibitem[\protect\citeauthoryear{{Cappelluti}, {Allevato} \&
  {Finoguenov}}{{Cappelluti} et~al.}{2012}]{Cappelluti2012}
{Cappelluti} N.,  {Allevato} V.,    {Finoguenov} A.,  2012, Advances in
  Astronomy, 2012

\bibitem[\protect\citeauthoryear{{Cisternas}, {Jahnke}, {Inskip}, {Kartaltepe},
  {Koekemoer}, {Lisker}, {Robaina}, {Scodeggio} \& {Sheth} K.}{{Cisternas}
  et~al.}{2011}]{Cisternas2011}
{Cisternas} M.,  {Jahnke} K.,  {Inskip} K.~J.,  {Kartaltepe} J.,  {Koekemoer}
  A.~M.,  {Lisker} T.,  {Robaina} A.~R.,  {Scodeggio} M.,    {Sheth} K. e.,
  2011, \apj, 726, 57

\bibitem[\protect\citeauthoryear{{Cowie}, {Garmire}, {Bautz}, {Barger},
  {Brandt} \& {Hornschemeier}}{{Cowie} et~al.}{2002}]{Cowie2002}
{Cowie} L.~L.,  {Garmire} G.~P.,  {Bautz} M.~W.,  {Barger} A.~J.,  {Brandt}
  W.~N.,    {Hornschemeier} A.~E.,  2002, \apjl, 566, L5

\bibitem[\protect\citeauthoryear{{Dressler}}{{Dressler}}{1980}]{Dressler1980}
{Dressler} A.,  1980, \apj, 236, 351

\bibitem[\protect\citeauthoryear{{Ebeling}, {Barrett}, {Donovan}, {Ma}, {Edge}
  \& {van Speybroeck}}{{Ebeling} et~al.}{2007}]{Ebeling2007}
{Ebeling} H.,  {Barrett} E.,  {Donovan} D.,  {Ma} C.-J.,  {Edge} A.~C.,    {van
  Speybroeck} L.,  2007, \apjl, 661, L33

\bibitem[\protect\citeauthoryear{{Ebeling}, {Edge}, {Bohringer}, {Allen},
  {Crawford}, {Fabian}, {Voges} \& {Huchra}}{{Ebeling}
  et~al.}{1998}]{Ebeling1998}
{Ebeling} H.,  {Edge} A.~C.,  {Bohringer} H.,  {Allen} S.~W.,  {Crawford}
  C.~S.,  {Fabian} A.~C.,  {Voges} W.,    {Huchra} J.~P.,  1998, \mnras, 301,
  881

\bibitem[\protect\citeauthoryear{{Ebeling}, {Edge}, {Mantz}, {Barrett},
  {Henry}, {Ma} \& {van Speybroeck}}{{Ebeling} et~al.}{2010}]{Ebeling2010}
{Ebeling} H.,  {Edge} A.~C.,  {Mantz} A.,  {Barrett} E.,  {Henry} J.~P.,  {Ma}
  C.~J.,    {van Speybroeck} L.,  2010, \mnras, 407, 83

\bibitem[\protect\citeauthoryear{{Ehlert}, {Allen}, {Brandt}, {Xue}, {Luo},
  {von der Linden}, {Mantz} \& {Morris}}{{Ehlert} et~al.}{2013}]{Ehlert2013}
{Ehlert} S.,  {Allen} S.~W.,  {Brandt} W.~N.,  {Xue} Y.~Q.,  {Luo} B.,  {von
  der Linden} A.,  {Mantz} A.,    {Morris} R.~G.,  2013, \mnras, 428, 3509

\bibitem[\protect\citeauthoryear{{Ehlert}, {von der Linden}, {Allen}, {Brandt},
  {Xue}, {Luo}, {Mantz}, {Morris}, {Applegate} \& {Kelly}}{{Ehlert}
  et~al.}{2014}]{EhlertFrac}
{Ehlert} S.,  {von der Linden} A.,  {Allen} S.~W.,  {Brandt} W.~N.,  {Xue}
  Y.~Q.,  {Luo} B.,  {Mantz} A.,  {Morris} R.~G.,  {Applegate} D.,    {Kelly}
  P.,  2014, \mnras, 437, 1942

\bibitem[\protect\citeauthoryear{{Elvis}, {Civano}, {Vignali}, {Puccetti} \&
  {Fiore} F.}{{Elvis} et~al.}{2009}]{Elvis2009}
{Elvis} M.,  {Civano} F.,  {Vignali} C.,  {Puccetti} S.,    {Fiore} F. e.,
  2009, \apjs, 184, 158

\bibitem[\protect\citeauthoryear{{Fan}, {Fang}, {Chen}, {Li}, {Lv}, {Kraiberg
  Knudsen} \& {Kong}}{{Fan} et~al.}{2014}]{Fan2014}
{Fan} L.,  {Fang} G.,  {Chen} Y.,  {Li} J.,  {Lv} X.,  {Kraiberg Knudsen} K.,
   {Kong} X.,  2014, ArXiv e-prints

\bibitem[\protect\citeauthoryear{{Ferrarese} \& {Merritt}}{{Ferrarese} \&
  {Merritt}}{2000}]{Ferrarese2000}
{Ferrarese} L.,  {Merritt} D.,  2000, \apjl, 539, L9

\bibitem[\protect\citeauthoryear{{Gabor}, {Impey}, {Jahnke}, {Simmons},
  {Trump}, {Koekemoer}, {Brusa}, {Cappelluti}, {Schinnerer}, {Smol{\v
  c}i{\'c}}, {Salvato}, {Rhodes}, {Mobasher}, {Capak}, {Massey}, {Leauthaud} \&
  {Scoville}}{{Gabor} et~al.}{2009}]{Gabor2009}
{Gabor} J.~M.,  {Impey} C.~D.,  {Jahnke} K.,  {Simmons} B.~D.,  {Trump} J.~R.,
  {Koekemoer} A.~M.,  {Brusa} M.,  {Cappelluti} N.,  {Schinnerer} E.,  {Smol{\v
  c}i{\'c}} V.,  {Salvato} M.,  {Rhodes} J.~D.,  {Mobasher} B.,  {Capak} P.,
  {Massey} R.,  {Leauthaud} A.,    {Scoville} N.,  2009, \apj, 691, 705

\bibitem[\protect\citeauthoryear{{Gehrels}}{{Gehrels}}{1986}]{Gehrels1986}
{Gehrels} N.,  1986, \apj, 303, 336

\bibitem[\protect\citeauthoryear{{Georgakakis}, {Coil}, {Laird}, {Griffith},
  {Nandra}, {Lotz}, {Pierce}, {Cooper}, {Newman} \& {Koekemoer}}{{Georgakakis}
  et~al.}{2009}]{Georgakakis2009}
{Georgakakis} A.,  {Coil} A.~L.,  {Laird} E.~S.,  {Griffith} R.~L.,  {Nandra}
  K.,  {Lotz} J.~M.,  {Pierce} C.~M.,  {Cooper} M.~C.,  {Newman} J.~A.,
  {Koekemoer} A.~M.,  2009, \mnras, 397, 623

\bibitem[\protect\citeauthoryear{{Gilmour}, {Best} \& {Almaini}}{{Gilmour}
  et~al.}{2009}]{Gilmour2009}
{Gilmour} R.,  {Best} P.,    {Almaini} O.,  2009, \mnras, 392, 1509

\bibitem[\protect\citeauthoryear{{Gunn} \& {Gott} III}{{Gunn} \&
  {Gott}}{1972}]{Gunn1972}
{Gunn} J.~E.,  {Gott} III J.~R.,  1972, \apj, 176, 1

\bibitem[\protect\citeauthoryear{{Haggard}, {Green}, {Anderson}, {Constantin},
  {Aldcroft}, {Kim} \& {Barkhouse}}{{Haggard} et~al.}{2010}]{Haggard2010}
{Haggard} D.,  {Green} P.~J.,  {Anderson} S.~F.,  {Constantin} A.,  {Aldcroft}
  T.~L.,  {Kim} D.-W.,    {Barkhouse} W.~A.,  2010, \apj, 723, 1447

\bibitem[\protect\citeauthoryear{{Hasinger}}{{Hasinger}}{2008}]{Hasinger2008}
{Hasinger} G.,  2008, \aap, 490, 905

\bibitem[\protect\citeauthoryear{{Hasinger}, {Miyaji} \& {Schmidt}}{{Hasinger}
  et~al.}{2005}]{Hasinger2005}
{Hasinger} G.,  {Miyaji} T.,    {Schmidt} M.,  2005, \aap, 441, 417

\bibitem[\protect\citeauthoryear{{Hopkins} \& {Beacom}}{{Hopkins} \&
  {Beacom}}{2006}]{Hopkins2006}
{Hopkins} A.~M.,  {Beacom} J.~F.,  2006, \apj, 651, 142

\bibitem[\protect\citeauthoryear{{Hopkins}, {Hernquist}, {Cox} \& {Kere{\v
  s}}}{{Hopkins} et~al.}{2008}]{Hopkins2008}
{Hopkins} P.~F.,  {Hernquist} L.,  {Cox} T.~J.,    {Kere{\v s}} D.,  2008,
  \apjs, 175, 356

\bibitem[\protect\citeauthoryear{{Kelly}, {von der Linden}, {Applegate},
  {Allen}, {Allen}, {Burchat}, {Burke}, {Ebeling}, {Capak}, {Czoske},
  {Donovan}, {Mantz} \& {Morris}}{{Kelly} et~al.}{2014}]{Kelly2012}
{Kelly} P.~L.,  {von der Linden} A.,  {Applegate} D.~E.,  {Allen} M.~T.,
  {Allen} S.~W.,  {Burchat} P.~R.,  {Burke} D.~L.,  {Ebeling} H.,  {Capak} P.,
  {Czoske} O.,  {Donovan} D.,  {Mantz} A.,    {Morris} R.~G.,  2014, \mnras,
  439, 28

\bibitem[\protect\citeauthoryear{{Kocevski}, {Faber}, {Mozena}, {Koekemoer},
  {Nandra}, {Rangel}, {Laird}, {Brusa}, {Wuyts} \& {Trump} J.~R.}{{Kocevski}
  et~al.}{2012}]{Kocevski2012}
{Kocevski} D.~D.,  {Faber} S.~M.,  {Mozena} M.,  {Koekemoer} A.~M.,  {Nandra}
  K.,  {Rangel} C.,  {Laird} E.~S.,  {Brusa} M.,  {Wuyts} S.,    {Trump} J.~R.
  e.,  2012, \apj, 744, 148

\bibitem[\protect\citeauthoryear{{Lehmer}, {Lucy}, {Alexander}, {Best},
  {Geach}, {Harrison}, {Hornschemeier}, {Matsuda}, {Mullaney}, {Smail},
  {Sobral} \& {Swinbank}}{{Lehmer} et~al.}{2013}]{Lehmer2013}
{Lehmer} B.~D.,  {Lucy} A.~B.,  {Alexander} D.~M.,  {Best} P.~N.,  {Geach}
  J.~E.,  {Harrison} C.~M.,  {Hornschemeier} A.~E.,  {Matsuda} Y.,  {Mullaney}
  J.~R.,  {Smail} I.,  {Sobral} D.,    {Swinbank} A.~M.,  2013, \apj, 765, 87

\bibitem[\protect\citeauthoryear{{Lehmer}, {Xue}, {Brandt}, {Alexander},
  {Bauer}, {Brusa}, {Comastri}, {Gilli}, {Hornschemeier}, {Luo}, {Paolillo},
  {Ptak}, {Shemmer}, {Schneider}, {Tozzi} \& {Vignali}}{{Lehmer}
  et~al.}{2012}]{Lehmer2012}
{Lehmer} B.~D.,  {Xue} Y.~Q.,  {Brandt} W.~N.,  {Alexander} D.~M.,  {Bauer}
  F.~E.,  {Brusa} M.,  {Comastri} A.,  {Gilli} R.,  {Hornschemeier} A.~E.,
  {Luo} B.,  {Paolillo} M.,  {Ptak} A.,  {Shemmer} O.,  {Schneider} D.~P.,
  {Tozzi} P.,    {Vignali} C.,  2012, \apj, 752, 46

\bibitem[\protect\citeauthoryear{{Mamon}}{{Mamon}}{1992}]{Mamon1992}
{Mamon} G.~A.,  1992, \apjl, 401, L3

\bibitem[\protect\citeauthoryear{{Mantz}, {Allen}, {Ebeling} \&
  {Rapetti}}{{Mantz} et~al.}{2008}]{Mantz2008}
{Mantz} A.,  {Allen} S.~W.,  {Ebeling} H.,    {Rapetti} D.,  2008, \mnras, 387,
  1179

\bibitem[\protect\citeauthoryear{{Mantz}, {Allen}, {Ebeling}, {Rapetti} \&
  {Drlica-Wagner}}{{Mantz} et~al.}{2010a}]{Mantz2010a}
{Mantz} A.,  {Allen} S.~W.,  {Ebeling} H.,  {Rapetti} D.,    {Drlica-Wagner}
  A.,  2010, \mnras, 406, 1773

\bibitem[\protect\citeauthoryear{{Mantz}, {Allen}, {Rapetti} \&
  {Ebeling}}{{Mantz} et~al.}{2010b}]{Mantz2010b}
{Mantz} A.,  {Allen} S.~W.,  {Rapetti} D.,    {Ebeling} H.,  2010, \mnras, 406,
  1759

\bibitem[\protect\citeauthoryear{{Martini}, {Miller}, {Brodwin}, {Stanford},
  {Gonzalez}, {Bautz}, {Hickox}, {Stern}, {Eisenhardt}, {Galametz}, {Norman},
  {Jannuzi}, {Dey}, {Murray}, {Jones} \& {Brown}}{{Martini}
  et~al.}{2013}]{Martini2013}
{Martini} P.,  {Miller} E.~D.,  {Brodwin} M.,  {Stanford} S.~A.,  {Gonzalez}
  A.~H.,  {Bautz} M.,  {Hickox} R.~C.,  {Stern} D.,  {Eisenhardt} P.~R.,
  {Galametz} A.,  {Norman} D.,  {Jannuzi} B.~T.,  {Dey} A.,  {Murray} S.,
  {Jones} C.,    {Brown} M.~J.~I.,  2013, \apj, 768, 1

\bibitem[\protect\citeauthoryear{{Moore}, {Lake} \& {Katz}}{{Moore}
  et~al.}{1998}]{Moore1998}
{Moore} B.,  {Lake} G.,    {Katz} N.,  1998, \apj, 495, 139

\bibitem[\protect\citeauthoryear{{Moretti}, {Campana}, {Lazzati} \&
  {Tagliaferri}}{{Moretti} et~al.}{2003}]{Moretti2003}
{Moretti} A.,  {Campana} S.,  {Lazzati} D.,    {Tagliaferri} G.,  2003, \apj,
  588, 696

\bibitem[\protect\citeauthoryear{{Popesso}, {Biviano}, {B{\"o}hringer} \&
  {Romaniello}}{{Popesso} et~al.}{2007}]{Popesso2007}
{Popesso} P.,  {Biviano} A.,  {B{\"o}hringer} H.,    {Romaniello} M.,  2007,
  \aap, 464, 451

\bibitem[\protect\citeauthoryear{{Puccetti}, {Vignali}, {Cappelluti}, {Fiore},
  {Zamorani}, {Aldcroft} \& {Elvis} M.}{{Puccetti} et~al.}{2009}]{Puccetti2009}
{Puccetti} S.,  {Vignali} C.,  {Cappelluti} N.,  {Fiore} F.,  {Zamorani} G.,
  {Aldcroft} T.~L.,    {Elvis} M. e.,  2009, \apjs, 185, 586

\bibitem[\protect\citeauthoryear{{Refregier} \& {Loeb}}{{Refregier} \&
  {Loeb}}{1997}]{Refregier1997}
{Refregier} A.,  {Loeb} A.,  1997, \apj, 478, 476

\bibitem[\protect\citeauthoryear{{Reichard}, {Heckman}, {Rudnick},
  {Brinchmann}, {Kauffmann} \& {Wild}}{{Reichard} et~al.}{2009}]{Reichard2009}
{Reichard} T.~A.,  {Heckman} T.~M.,  {Rudnick} G.,  {Brinchmann} J.,
  {Kauffmann} G.,    {Wild} V.,  2009, \apj, 691, 1005

\bibitem[\protect\citeauthoryear{{Sarazin}}{{Sarazin}}{1988}]{Sarazin1988}
{Sarazin} C.~L.,  1988, {X-ray emission from clusters of galaxies}.
Cambridge Astrophysics Series, Cambridge: Cambridge University Press

\bibitem[\protect\citeauthoryear{{Schawinski}, {Simmons}, {Urry}, {Treister} \&
  {Glikman}}{{Schawinski} et~al.}{2012}]{Schawinski2012}
{Schawinski} K.,  {Simmons} B.~D.,  {Urry} C.~M.,  {Treister} E.,    {Glikman}
  E.,  2012, \mnras, 425, L61

\bibitem[\protect\citeauthoryear{{Schawinski}, {Treister}, {Urry}, {Cardamone},
  {Simmons} \& {Yi}}{{Schawinski} et~al.}{2011}]{Schawinski2011}
{Schawinski} K.,  {Treister} E.,  {Urry} C.~M.,  {Cardamone} C.~N.,  {Simmons}
  B.,    {Yi} S.~K.,  2011, \apjl, 727, L31

\bibitem[\protect\citeauthoryear{{Silk} \& {Rees}}{{Silk} \&
  {Rees}}{1998}]{Silk1998}
{Silk} J.,  {Rees} M.~J.,  1998, \aap, 331, L1

\bibitem[\protect\citeauthoryear{{Simmons}, {Urry}, {Schawinski}, {Cardamone}
  \& {Glikman}}{{Simmons} et~al.}{2012}]{Simmons2012}
{Simmons} B.~D.,  {Urry} C.~M.,  {Schawinski} K.,  {Cardamone} C.,    {Glikman}
  E.,  2012, \apj, 761, 75

\bibitem[\protect\citeauthoryear{{Tal}, {van Dokkum}, {Nelan} \&
  {Bezanson}}{{Tal} et~al.}{2009}]{Tal2009}
{Tal} T.,  {van Dokkum} P.~G.,  {Nelan} J.,    {Bezanson} R.,  2009, \aj, 138,
  1417

\bibitem[\protect\citeauthoryear{{Truemper}}{{Truemper}}{1993}]{Trumper1993}
{Truemper} J.,  1993, Science, 260, 1769

\bibitem[\protect\citeauthoryear{{Ueda}, {Akiyama}, {Hasinger}, {Miyaji} \&
  {Watson}}{{Ueda} et~al.}{2014}]{Ueda2014}
{Ueda} Y.,  {Akiyama} M.,  {Hasinger} G.,  {Miyaji} T.,    {Watson} M.~G.,
  2014, ArXiv e-prints

\bibitem[\protect\citeauthoryear{{Ueda}, {Akiyama}, {Ohta} \& {Miyaji}}{{Ueda}
  et~al.}{2003}]{Ueda2003}
{Ueda} Y.,  {Akiyama} M.,  {Ohta} K.,    {Miyaji} T.,  2003, \apj, 598, 886

\bibitem[\protect\citeauthoryear{{van Dokkum}}{{van
  Dokkum}}{2001}]{VanDokkum2001}
{van Dokkum} P.~G.,  2001, \pasp, 113, 1420

\bibitem[\protect\citeauthoryear{{Vikhlinin}, {Kravtsov}, {Burenin}, {Ebeling},
  {Forman}, {Hornstrup}, {Jones}, {Murray}, {Nagai}, {Quintana} \&
  {Voevodkin}}{{Vikhlinin} et~al.}{2009}]{Vikhlinin2009}
{Vikhlinin} A.,  {Kravtsov} A.~V.,  {Burenin} R.~A.,  {Ebeling} H.,  {Forman}
  W.~R.,  {Hornstrup} A.,  {Jones} C.,  {Murray} S.~S.,  {Nagai} D.,
  {Quintana} H.,    {Voevodkin} A.,  2009, \apj, 692, 1060

\bibitem[\protect\citeauthoryear{{von der Linden}, {Allen}, {Applegate},
  {Kelly}, {Allen}, {Ebeling}, {Burchat}, {Burke}, {Donovan}, {Morris},
  {Blandford}, {Erben} \& {Mantz}}{{von der Linden}
  et~al.}{2014}]{VonderLinden2012}
{von der Linden} A.,  {Allen} M.~T.,  {Applegate} D.~E.,  {Kelly} P.~L.,
  {Allen} S.~W.,  {Ebeling} H.,  {Burchat} P.~R.,  {Burke} D.~L.,  {Donovan}
  D.,  {Morris} R.~G.,  {Blandford} R.,  {Erben} T.,    {Mantz} A.,  2014,
  \mnras, 439, 2

\bibitem[\protect\citeauthoryear{{Xue}, {Brandt}, {Luo}, {Rafferty},
  {Alexander}, {Bauer}, {Lehmer}, {Schneider} \& {Silverman}}{{Xue}
  et~al.}{2010}]{Xue2010}
{Xue} Y.~Q.,  {Brandt} W.~N.,  {Luo} B.,  {Rafferty} D.~A.,  {Alexander} D.~M.,
   {Bauer} F.~E.,  {Lehmer} B.~D.,  {Schneider} D.~P.,    {Silverman} J.~D.,
  2010, \apj, 720, 368

\bibitem[\protect\citeauthoryear{{Xue}, {Luo}, {Brandt}, {Bauer}, {Lehmer},
  {Broos}, {Schneider}, {Alexander}, {Brusa}, {Comastri}, {Fabian}, {Gilli} \&
  {Hasinger}}{{Xue} et~al.}{2011}]{Xue2011}
{Xue} Y.~Q.,  {Luo} B.,  {Brandt} W.~N.,  {Bauer} F.~E.,  {Lehmer} B.~D.,
  {Broos} P.~S.,  {Schneider} D.~P.,  {Alexander} D.~M.,  {Brusa} M.,
  {Comastri} A.,  {Fabian} A.~C.,  {Gilli} R.,    {Hasinger} G.,  2011, \apjs,
  195, 10

\end{thebibliography}

\end{document}